 \newcommand\aap{{\em A\&A}}
 \newcommand\aaps{{\em A\&AS}}
 \newcommand\aj{{\em AJ}}
 \newcommand\apj{{\em ApJ}}
 \newcommand\apjl{{\em ApJ}}
 \newcommand\apjs{{\em ApJS}}

 \newcommand\mnras{{\em MNRAS}}

 \newcommand\pasp{{\em PASP}}
 \documentclass[usenatbib]{mn2e} 

 \usepackage[dvips]{graphicx}
 \usepackage{color}
\usepackage{setspace}
 
\begin{document}

 %----------------------------------------------------

 \title[Low X-ray Luminosity Galaxy Clusters ]
 {Low X-Ray Luminosity Galaxy Clusters. II.  
 Optical properties and morphological content at 0.18 $<$ z $<$ 0.70\thanks {Based on observations obtained at the Gemini Observatory, which is 
 operated by Association of Universities for Research in Astronomy, Inc., under 
 cooperative agreement with the NSF on behalf of the Gemini partnership: the 
 National Science Foundation (United State), the Particle Physics and Astronomy 
 Research Council (Canada), CONICYT (Chile), the Australian Research Council 
 (Australia), CNPq (Brazil) and CONICET (Argentina)
 {\dag} : jnilo@mail.oac.uncor.edu, jnilo@dfuls.cl}
 }
 \author[Nilo Castell\'on et al.]
 {
 \parbox[t]{\textwidth}{
 Jos\'e Luis Nilo Castell\'on$^{1,2}{\dag}$; 
 M. Victoria Alonso$^{1,3}$; Diego Garc\'ia Lambas$^{1,3}$; 
Ana Laura O' Mill$^{4}$; Carlos Valotto$^{1,3}$; E. Rodrigo Carrasco$^{5}$; 
H\'ector Cuevas$^{2}$ \& Amelia Ram\'irez$^{2}$}
 \vspace*{6pt} \\
 $^1$ Instituto de Astronom\'ia Te\'orica y Experimental, (IATE-CONICET), Laprida 922, C\'ordoba, Argentina.\\
 $^2$ Departamento de F\'isica, Facultad de Ciencias, Universidad de La Serena, Cisternas 1200, La Serena, Chile. \\
 $^3$ Observatorio Astron\'omico de C\'ordoba, Universidad Nacional de C\'ordoba, Laprida 854, C\'ordoba, Argentina. \\
$^4$ Instituto de Astronomia, Geof\'isica e Ci\^encias Atmosf\'ericas da USP, 
Rua do Mat\~ao 1226, Cidade Universit\'aria, 05508-090, S\~ao Paulo, Brazil.\\
 $^5$ Gemini Observatory/AURA, Southern Operations Center, Casilla 603, La Serena, Chile}

 \date{\today}

 \pagerange{\pageref{firstpage}--\pageref{lastpage}}

 \maketitle

 \label{firstpage}

 \begin{abstract}

 This is the second of a series of papers on low X-ray luminosity galaxy 
clusters, in which 
we present the $r^\prime$, $g^\prime$ and $i^\prime$ photometry obtained with 
GMOS-IMAGE at Gemini North and South telescopes for seven systems in the 
redshift range of 0.18 to 0.70. Optical magnitudes, colours and 
morphological parameters, namely, concentration index, ellipticity and  
visual morphological classification, are also given.

At lower redshifts, the presence of a well-defined red cluster sequence 
extending by more than 4 magnitudes showed that these intermediate-mass 
clusters 
had reached a relaxed stage. This was confirmed by the small fraction of 
blue galaxy members observed in the central regions of $\sim$ 0.75 Mpc.  
In contrast, galaxy clusters at higher redshifts had a 
less important red cluster sequence.  We also found that the 
galaxy radial density profiles in these clusters were well fitted by a 
single power law.

At 0.18 $<$ z $<$ 0.70, we observed an increasing fraction of blue galaxies and 
a decreasing fraction of 
lenticulars, with the early-type fraction remaining almost constant. Overall, 
the 
results of these intermediate-mass clusters are in agreement with those for 
high mass clusters.

 \end{abstract}

 \begin{keywords}
 galaxies: clusters: general, galaxies: photometry, galaxies: fundamental
parameters.
 \end{keywords}

 \section{INTRODUCTION}

 Clusters of galaxies
 are ideal systems to investigate the assembly of the structures in the 
Universe. 
 In massive clusters, Dressler et al. (1997) found significant changes in 
 their morphological content as a function of redshift, with these results 
 being explained by the major episodes of star formation produced in the 
galaxies at higher 
 redshifts followed by a subsequent passive evolution with  
 little star formation activity (Ellis et al. 1997; Gladders et al. 1998;
 Stanford et al. 1998). However,  Poggianti et al. (2009) showed that at 
 redshifts of 0.5 $<$ z $<$ 1.2, the
 morphological evolution was not only restricted to massive clusters. In
 lower mass systems, the fraction of spiral and S0 galaxies evolved more 
 strongly while ellipticals remained unchanged.

 Colour-Magnitude
 Relations (CMRs) were first used to find galaxy clusters and to 
estimate their
 redshifts (Yee, Gladders \& L\'opez-Cruz, 1999).  
 Since then, the slope and
 scatter of this relation have been studied for different cluster samples with diverse
 masses and redshifts. For instance,
changes in the CMR slope with redshift were able to constrain the formation 
epoch 
of galaxy clusters (Gladders et al. 1998). Mei et al. (2009)
 studied the CMR scatter to obtain the average age of early-type galaxies
 in the clusters, but found no 
 significant dependence on cluster mass or redshift 
 (e.g., Kodama \& Arimoto 1997; Kauffman \& Charlot 1998;
 Bernardi et al. 2005; Gallazzi et al. 2006; Tran et al. 2007).
 
 The Red Cluster Sequence (hereafter, RCS) is a tight sequence in the 
 colour-magnitude plane defined by early-type galaxies 
 (Visvanathan \& Sandage 1977; 
 Gladders et al. 1998; De Lucia et al. 2004; Gilbank et al. 2008; 
 Lerchster et al. 2011), which was used by Mullis et al. (2005) as a 
 reliable distance indicator of massive, X-ray-luminous galaxy clusters 
 at z $\sim$ 1.4. In addition, Demarco et al. (2010) confirmed 
 some red-sequence galaxy clusters spectroscopically at redshifts higher than 1.

 Low X-ray luminosity (ie. intermediate-mass) clusters have not been 
extensively studied 
compared to massive, luminous X-ray systems.  
 Among these existing few studies, Balogh et al. (2002) found that in 
intermediate 
 X-ray luminosity  clusters at  0.23 $<$ z $<$  0.3, the galaxy 
properties are similar to those in massive clusters at the same redshift 
range. Moreover,
Jeltema et al. (2006) found scaling relations between luminosity, 
temperature, and velocity dispersion 
 in six 
 groups at 0.2 $<$ z $<$  0.6, which were comparable to those observed in 
nearby galaxy systems.
 Carrasco et al. (2007) studied the low luminosity X-ray cluster RX J1117.4+0743
  at z=0.485 and reported a complex morphology existing composed of at least 
two substructures in velocity space.  This cluster also presented an offset 
between 
 the position of the brightest group galaxy and the X-ray emission centre.

More recently, Balogh et al. (2011) conducted a morphological study of 
similar clusters at 0.85 $<$ z $<$ 1, providing more evidence that
group environment plays an important role in galaxy evolution.
Based on colours, 
a prominent transient population was identified which moved 
from the blue cloud to the red sequence.  In addition, Connelly et al. (2012) 
presented an extensive spectroscopic study of
X-ray and optically defined systems at intermediate redshifts, which contributed
to the understanding of galaxy evolution, in particular through the 
L$_X$--$\sigma$ relation of these systems.  

 Our aim is to shed new light on the cluster 
 assembly of low X-ray luminosity galaxy clusters within
 the hierarchical formation scenario.  The first paper of the series 
 (Nilo Castell\'on et al. 2013, hereafter Paper I) contained the main goals, 
sample
 selection, and details of observations and data
 reduction for both photometry and spectroscopy.
 Membership was defined as galaxies
 with projected distances of less than 0.75 Mpc from
 the cluster centre. For galaxy clusters observed with spectroscopy, membership
 was also restricted to objects having
 differences in radial velocities ($\Delta$V) relative to the mean cluster 
 smaller than  the cluster velocity dispersion.  
For those clusters without any available spectroscopy, we used the studies of  
O' Mill et al. 
(2012) and Aihara et al. (2011) to estimate the photometric redshifts for 
 all objects in the cluster fields. Following Paper I, taking into account higher
 uncertainties in the photometric redshift estimates, members had $\Delta$V 
values of approximately 6000 km~s$^{-1}$.

 Here, in this second paper, we present the photometric 
 properties of seven low X-ray luminosity, 
 intermediate-mass galaxy clusters observed with Gemini telescopes.
 In $\S$2, we present the photometric data and morphological
 parameters of the cluster members, together with an example of the 
catalogue. In $\S$3,  we analyse photometric properties namely galaxy number 
counts, Colour-Magnitude
 relations with the red cluster sequence and colours, galaxy number density 
 profiles, galaxy projected 
 distributions and morphological content.
 In $\S$4, we present a summary of the main results for these seven
 intermediate-mass galaxy clusters. 

 For all cosmology-dependent calculations, we assume $\Omega_\Lambda$=0.7, $\Omega_m $=0.3 and $h$=0.7.

 \section{Photometric Data}

 We selected nineteen galaxy clusters from the ROSAT PSPC 
 Pointed Observations 
 (Vikhlinin et al. 1998 and Mullis et al. 2003) for this project. 
 These had 
 X-ray luminosities in the range L$_X \sim$ 0.5--45  
 $\times$ $10^{43}$ erg s$^{-1}$,  angular core-radius smaller than 60 arcsec
 and redshifts between 0.15 to 0.70 (for further details see Paper I).
 The galaxy photometric properties of seven clusters observed with
 Gemini North and South telescopes using the Multi-Object Spectrograph 
 (Hook et al. 2004, hereafter GMOS) in the Image mode are presented. The 
detector is an array of three 2048 $\times$ 4608 pixels EEV CCDs, and using 
a 2$\times$2 
 binning, the pixel scale is 0.1454 arcsec per pixel which corresponds to a
 field of view (FOV) of 5.5 arcmin\ensuremath{^2} of the sky.  
This strategy 
 allowed us to study about 1.2 Mpc of the cluster centre, regardless of the 
redshift. 

 All galaxy clusters and Landolt (1992) standard stars were observed using
the $r^\prime$ filter; $g^\prime$ or $i^\prime$ filters were used to obtain 
colours. Table~\ref{table1} shows a summary of 
 the main characteristics of the clusters including the 
 identification from Vikhlinin et al. (1998); the equatorial 
 coordinates of the X-ray emission peak, the
 X-ray luminosity in the 0.5--2.0 keV energy band and also the mean redshift 
measured by Mullis et al. (2003)
 together with the Gemini Program identification and the observing passbands 
(including number 
 of exposures and individual exposure time in seconds, respectively).  
All images were observed 
 under good photometric conditions with mean seeing values of 0.75, 0.66 and 
0.74 arcsec in the $g^\prime$, $r^\prime$ and $i^\prime$ filters, respectively. 
The last observations (GN-2011A-Q-75) were taken under excellent conditions,
 with seeing in the $r^\prime$ band being between 0.48 and 0.55 arcsec. 

The colour composed images obtained in the central parts of these studied 
clusters can be found in Paper I, where 
 the standard reduction and galaxy detection are specified. 
 These images were processed with the v1.4 Gemini IRAF\footnote{IRAF is distributed by NOAO, which is 
 operated by the 
 Association of Universities for Research in Astronomy Inc., under cooperative 
 agreement with the National Science Foundation} package, where they were 
overscanned and 
 bias subtracted, as well as being trimmed and flat-fielded before being 
combined to 
 create the final images.  SExtractor v2.5.0 (Bertin \& Arnouts, 1996) was
 used for galaxy detection and to obtain the main photometric parameters.
 Using the star-galaxy separation, galaxies were defined as those 
 objects having an isophotal axis ratio b/a $>$ 0.9, CLASS\_STAR $<$ 0.8 
(a parameter associated
 with the light distribution) and half-light
  radius $>$ 5 arcsec. 
 The galaxy total magnitudes were of SExtractor Kron magnitude 
 (MAG\_AUTO) and the colours were aperture
 magnitudes obtained within a fixed aperture of 20 pixels (equivalent 
 to 2.9 arcsec).  The photometric procedure is extensively detailed in 
 Paper I.

 SExtractor overestimates the aperture magnitudes of the brightest galaxies in 
 clusters, an effect which is particularly important at lower redshifts. 
  For the bright objects in [VMF98]022 and [VMF98]124, we used aperture 
 magnitudes within an
 80 pixel radius as their total magnitudes.  Final magnitudes were expressed 
in the 
 AB system, after correcting for galactic extinction using reddening maps 
from Schlegel et al. (1998).  Magnitude limits and completeness levels for 
the galaxy clusters were 
 obtained through simulated catalogues and images using the procedures explained
 in Paper I.  Table~\ref{table2} shows the r$^{\prime}$ magnitudes and their 
Poisson errors obtained for 
50\%  and 90\% completeness levels in the studied fields.  The deepest images
corresponded to the best observation conditions mentioned above.

 \subsection {Morphological Parameters}

 One of our main goals was to study galaxy morphology evolution in low and intermediate-mass 
 clusters. As mentioned by many authors (Fasano et al. 
 2000, van den Bergh 2001, Vulcani et al. 2010, Calvi et al. 2011), a 
 correct morphology assignment to a galaxy depends on several factors such 
as the observed instrumental properties,  the passbands, the quality of the 
observations
 and galaxy redshifts, among others.  The quality of our observations allowed
 us to 
 resolve structures with sizes 
 between 1.5 to 4 kpc according to the galaxy redshifts, and therefore it was
possible to perform a
 morphological classification based on visual inspection. 
This classification was carried out
 by four of our team (JNC, MVA, HC, AR) using r$^\prime$ postage stamps with 
a fixed 
 size of 50$\times$50 kpc. No information concerning galaxy redshift, colours, 
cluster properties, or spectroscopic features was used.

 Five different morphological types were defined. Elliptical galaxies (T=1) were
 those objects having extended and bright halos with a clear 
 spheroidal morphology and without substructures or asymmetries. S0/Sa galaxies 
(T=2) had extended halos and bright but concentrated cores, and also 
 included in this type were those galaxies with a clear edge-on lenticular 
shape. Spiral 
 galaxies (T=3) were objects with clear arms, or/and an extended and tight 
 edge-on disc with bright cores. Irregulars (T=4) were objects having no clear 
 structures or possible mergers (T=5).

\begin{figure*}
 \includegraphics[width=30mm,height=30mm]{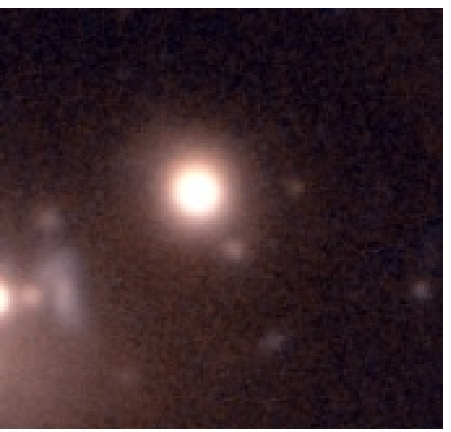} 
 \includegraphics[width=30mm,height=30mm]{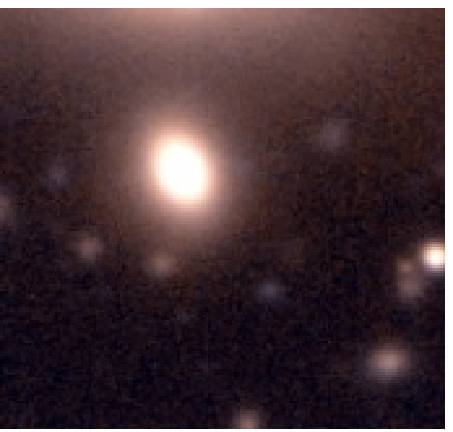} 
 \includegraphics[width=30mm,height=30mm]{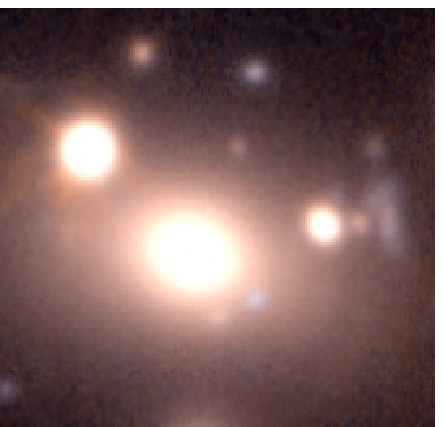} 
 \includegraphics[width=30mm,height=30mm]{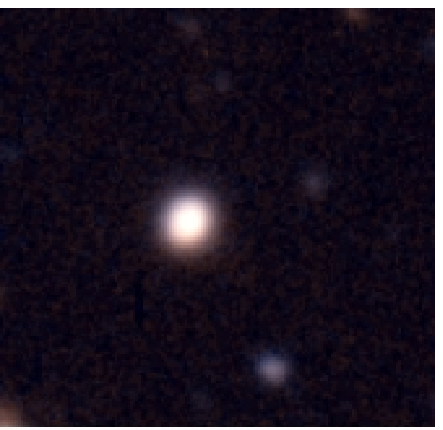}
 \includegraphics[width=30mm,height=30mm]{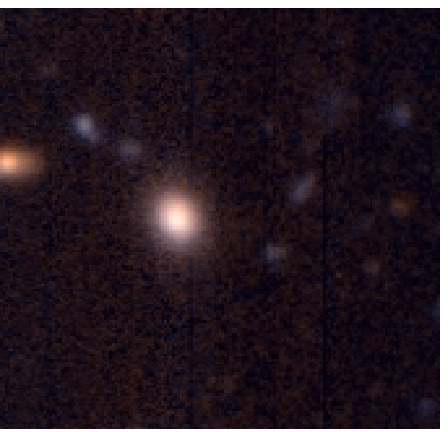} 
 \includegraphics[width=30mm,height=30mm]{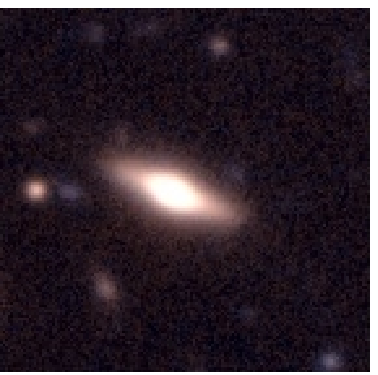}
 \includegraphics[width=30mm,height=30mm]{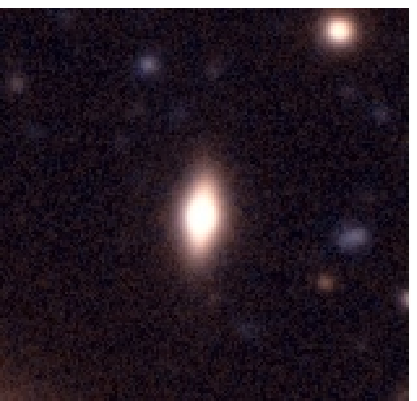} 
 \includegraphics[width=30mm,height=30mm]{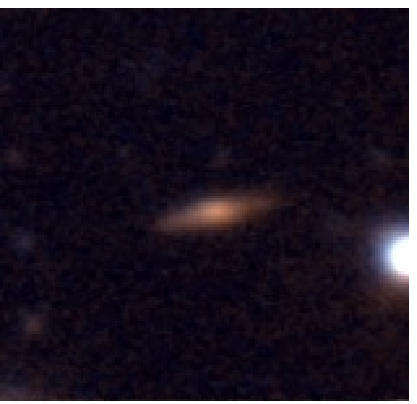}  
 \includegraphics[width=30mm,height=30mm]{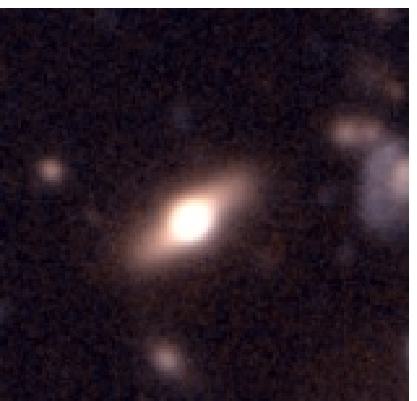} 
 \includegraphics[width=30mm,height=30mm]{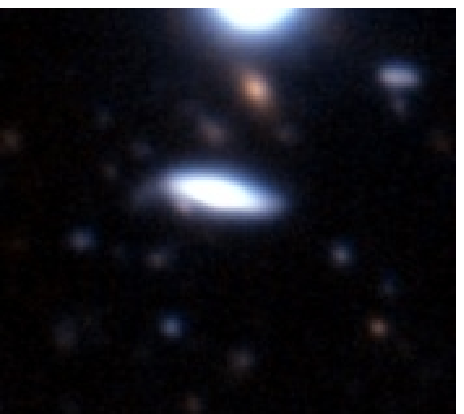} 
 \includegraphics[width=30mm,height=30mm]{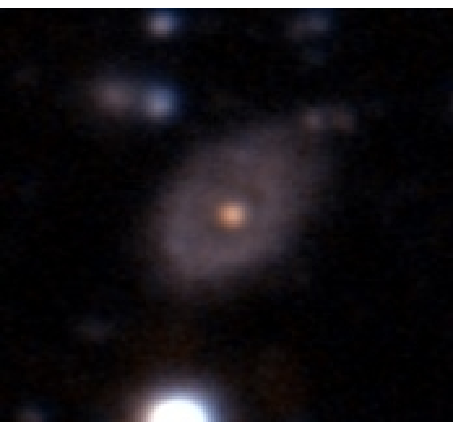} 
 \includegraphics[width=30mm,height=30mm]{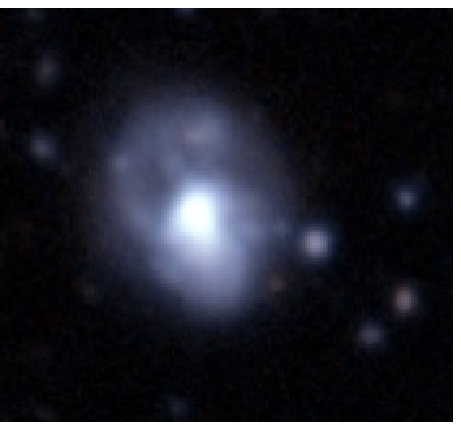} 
 \includegraphics[width=30mm,height=30mm]{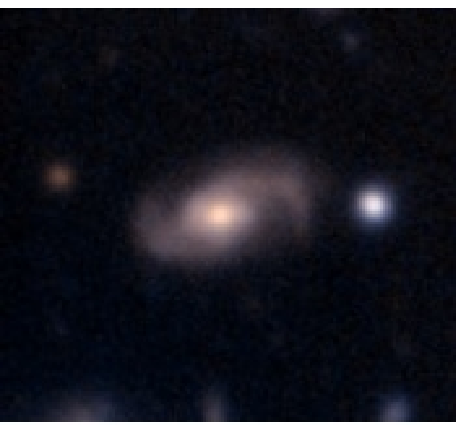}   
 \includegraphics[width=30mm,height=30mm]{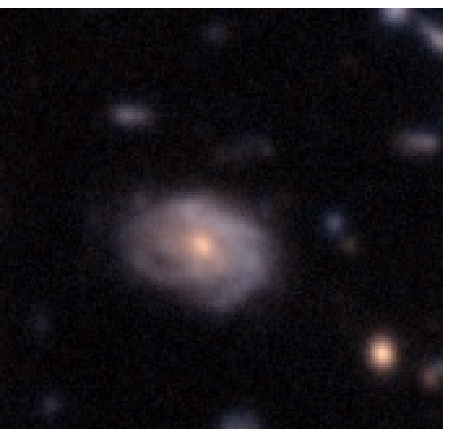} 
 \includegraphics[width=30mm,height=30mm]{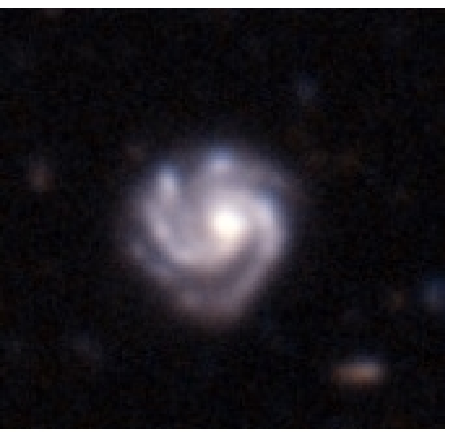} 
 \includegraphics[width=30mm,height=30mm]{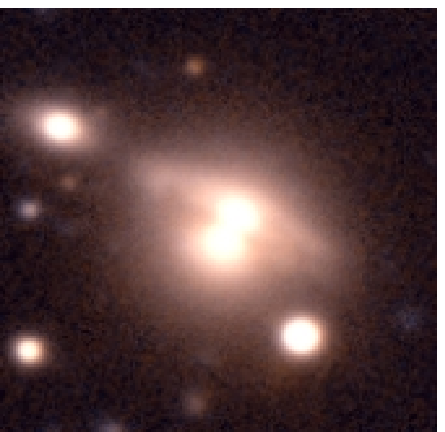} 
 \includegraphics[width=30mm,height=30mm]{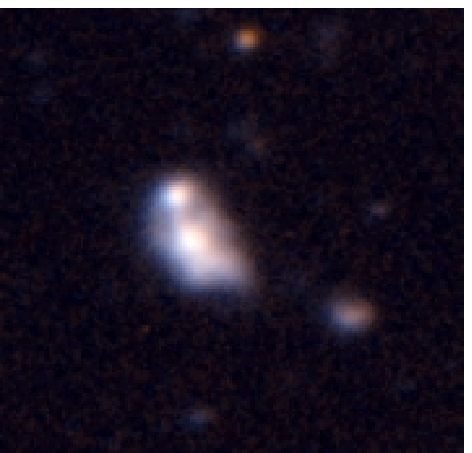} 
 \includegraphics[width=30mm,height=30mm]{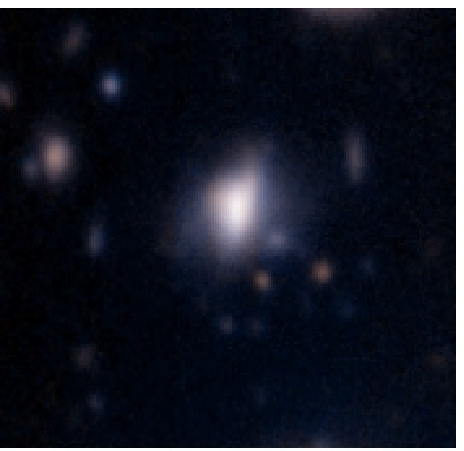} 
 \includegraphics[width=30mm,height=30mm]{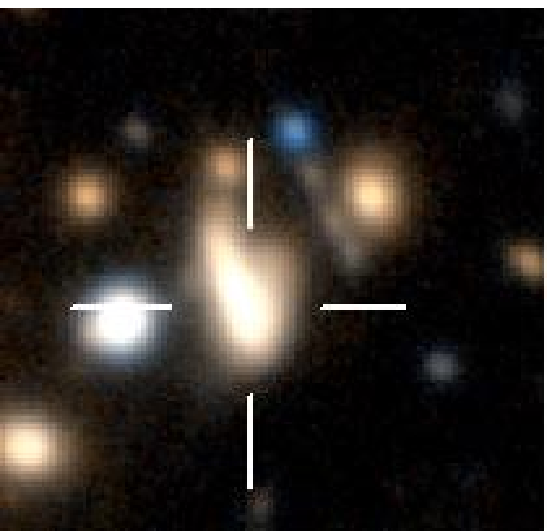}
  \includegraphics[width=30mm,height=30mm]{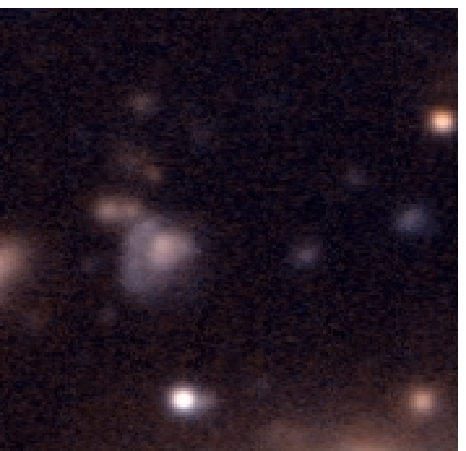}
 \caption{Morphological type classification.  Some examples of galaxies 
 classified in this paper, from top to bottom:  ellipticals, 
 lenticulars, spirals, 
 irregulars and mergers/peculiars.}
 \label{mtype}
 \end{figure*}

Consistently, for about 71\% of galaxies, 
the same morphological type was assigned by the different observers.  For 19\% of the cases, only two of the team
gave the same morphological type. To define the morphological type for those 
objects with no 
clear assignment, another
classification was performed by using the observed images in all the filters and the
combined image. This resulted in less than 2\% of the cases having no clear 
structure and being in 
general small-sized 
faint galaxies which were assigned T=4.  As an independent check on our 
procedure, we 
used the available spectra with good S/N for 75 galaxies, taking 
into account the continuum spectral shape and some characteristic emission 
and/or absorption lines.  The final assignment of galaxy type considered the 
spectral library of Kinney et al. (1996) finding a good agreement 
in 59 galaxies.  Some examples of our morphological 
classification are given in  
 Figure~\ref{mtype}.  From top to bottom, it is shown ellipticals, S0s, 
spirals and 
irregulars or peculiar galaxies. Synthetic images of 
100 galaxies with different redshifts and morphologies were created 
consistently with the observational conditions using STUFF and 
 Skymaker (explained in Paper I).  These synthetic galaxies were then 
classified using the same 
criteria defined above, with there being no coincident  morphology
assignment only for 10 faint objects. 

 The concentration index (C) is a suitable measurement of  
 galaxy structure, and due to it being a robust and easy way to measure 
objects automatically, it is ideal for large galaxy surveys (Conselice et al. 
2000, Strateva et al. 2001, Yamauchi et al. 2005). This 
 parameter has been used as the first indicator of the galaxy 
 morphology in various studies 
 (Hashimoto \& Oemler 1999; Yamauchi et al. 2005 and Conselice, 
Rajgor \& Myers 2008). However, using the Hubble Space Telescope GOODS ACS 
and Hubble Deep Field images, 
Conselice (2003) has shown that the C parameter
can only be reliably measured out to z $\sim$ 3. 

In this work, we used the concentration index as defined by the ratio of two
circular radii which contain 80 and 20 percent of the total Petrosian
flux. In this way, centrally concentrated ellipticals presented 
 larger C values compared with those of late type galaxies.
 For a classical de Vaucouleurs profile, C is approximately 5, while for an 
 exponential disk it has a value of about 2 (Conselice 2003).
 For our data, low values of C were found in general and presented a large 
 scatter for each morphological type.

\begin{figure*} 
 \includegraphics[width=60mm,height=60mm]{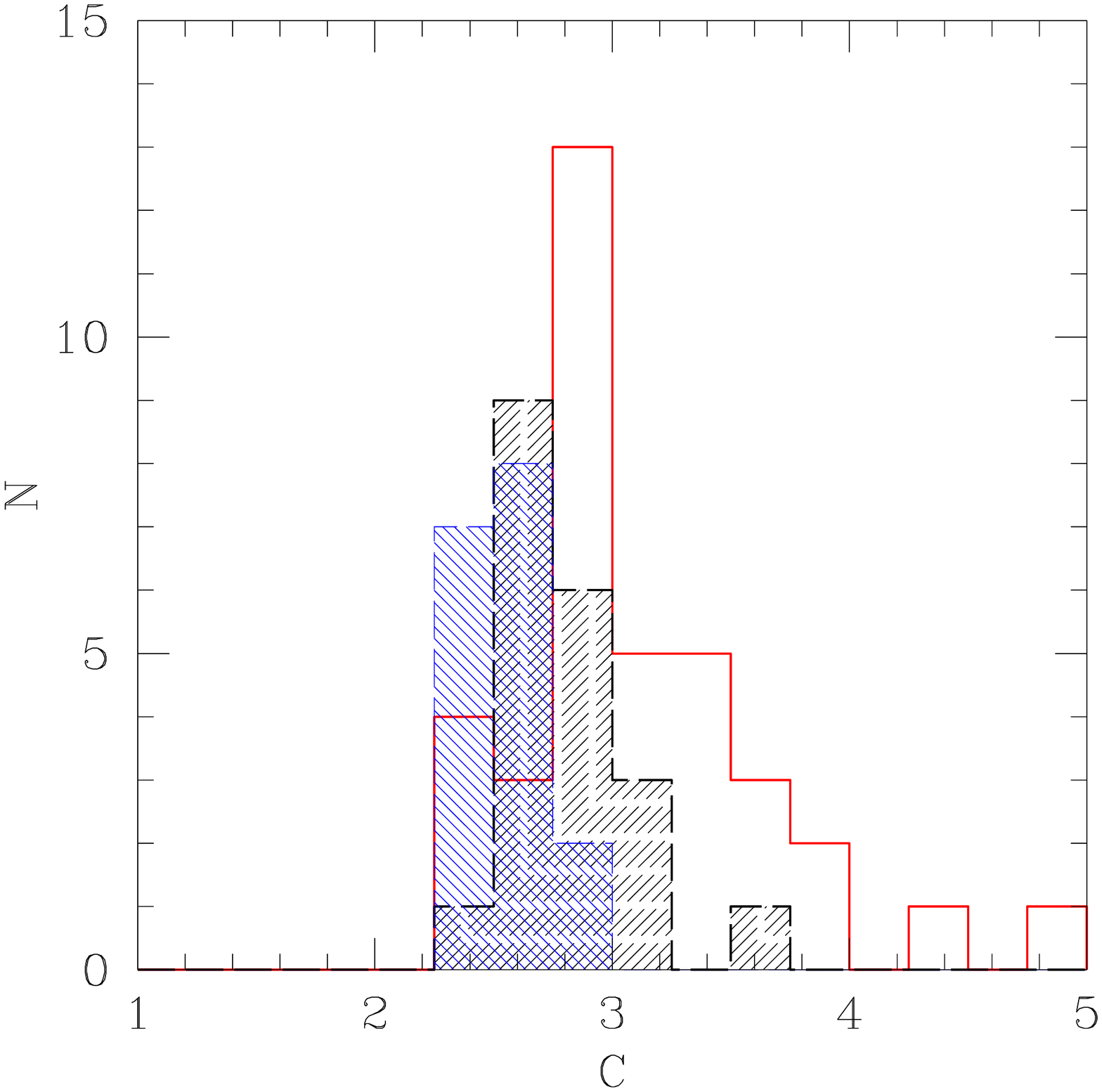}
 \includegraphics[width=60mm,height=60mm]{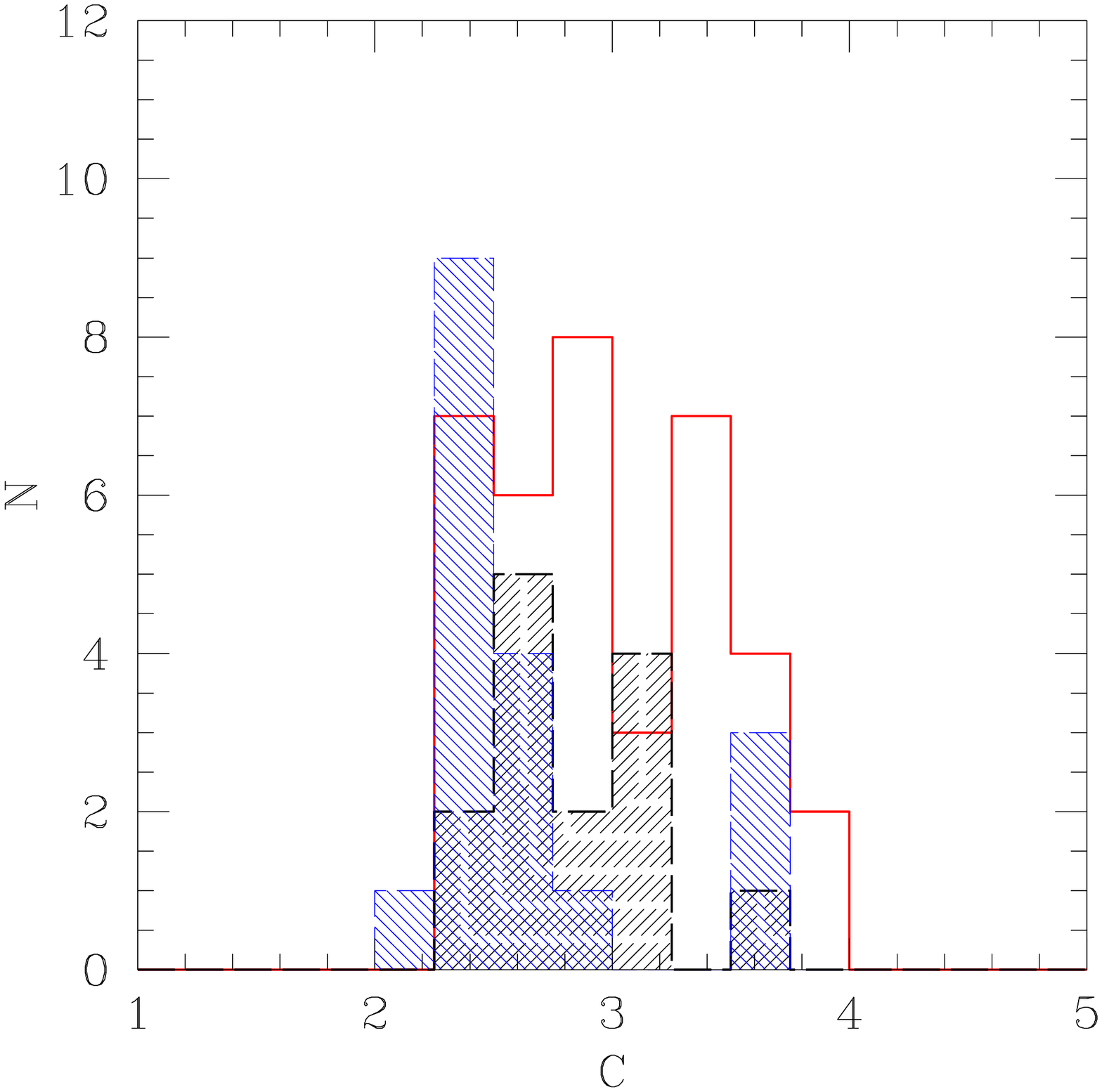}
 \caption{Concentration index distributions for all visually 
classified galaxies. Low (high) redshift clusters are shown in left (right) 
panels.  Histogram colours correspond to ellipticals (red), lenticulars 
(black shaded) and spirals (blue). }
 \label{histos}
 \end{figure*}

 Figure~\ref{histos} shows the concentration index distributions
for all galaxies visually classified as ellipticals, lenticulars and spirals.  
The left panel 
contains the objects of the three lower redshift clusters, while the right panel
has objects from clusters at higher redshifts.  
In general, at lower redshifts, the distributions were in agreement with 
the normal properties of the three
galaxy morphologies. The C 
distribution for early-type objects was well defined with a peak at about 2.8,
whereas for higher redshifts this distribution was broader with several peaks 
between 2.4 and 3.5.  After checking that galaxies with lower C values 
corresponded to 
objects with fainter
 r$^\prime$ total magnitudes, we were then able to use C, b/a and the visual 
morphological type T as suitable 
morphological 
 parameters to study the 
 photometric properties 
 of the galaxies in these low X-ray clusters.

 \subsection{The Final Catalogues}

As an example of the catalogues Table~\ref{table3} shows the
following photometric properties together with the redshift of the 26 member 
galaxies in the neighbourhoods of the
 cluster [VMF98]022: galaxy identification; J2000 equatorial 
 coordinates; $r^\prime$ total magnitude and aperture colour (within 2.9 
arcsec); structural parameters such as 
 concentration index, ellipticity, visual classification and redshift.
 The catalogues from the studied clusters can be accessed at 
http://www.dfuls.cl/XGClusters/.
 As mentioned before, the magnitudes were given in the AB system, after being
fully corrected by galactic extinction.

 \section{DATA ANALYSIS}

We restricted our analysis to galaxies within 0.75 Mpc from the cluster 
centre, a limit that is small enough to include the 
virialized region of the studied clusters. Although a larger radii 
should be used for richer systems, 
for reasons of simplicity we adopted this 
radial cut at 0.75 Mpc, which removed the presence of a significant bias across the 
cluster sample due to the different weightings of external and internal regions.
In this way, we were able to perform the same analysis on all the clusters regardless
of their redshifts.

To determine cluster centres, two different criteria were used: the centre of a 
dominant galaxy (the Bright Cluster Galaxy, hereafter BCG, at least
1.5 magnitudes brighter than the second 
ranked cluster galaxy) in the three low redshift clusters ([VMF98]001, 
[VMF98]093 and [VMF98]124) and the peak of the X-ray emission in the four 
high redshift clusters.  In the first case, the difference between the peak 
of the X-ray emission and the BCG centre was less than the X-ray peak 
position uncertainty.

In this section, we used the photometric properties of the galaxies defined as
 members in the seven studied galaxy clusters.  For galaxies with 
spectroscopic redshifts, cluster membership was assigned to galaxies with 
clustercentric 
radial velocity differences smaller than
 the velocity dispersion of the cluster.  However, for those galaxies with only
photometric redshift estimates, we considered as members those 
 objects with radial velocity relative differences less than 6000 km~s$^{-1}$.

 \subsection{Number Counts}

The r$^{\prime}$ number counts were defined as the number of galaxies in the 
 magnitude range of 16 to 27 mag in 0.5 magnitude bins per square degrees.
 Our results were compared with models from Nagashima et al. (2002) 
for three 
 different cosmologies: a standard cold dark matter universe (SC), a 
low-density 
 flat universe with a non-zero cosmological constant (LC) and a low density 
open 
 universe with zero cosmological constant (OC).  The photometric relations 
 were taken from Fukugita et al. (1995) and incorporated into the models after 
being transformed to our photometric system.

\begin{figure}
 \includegraphics[width=90mm,height=90mm]{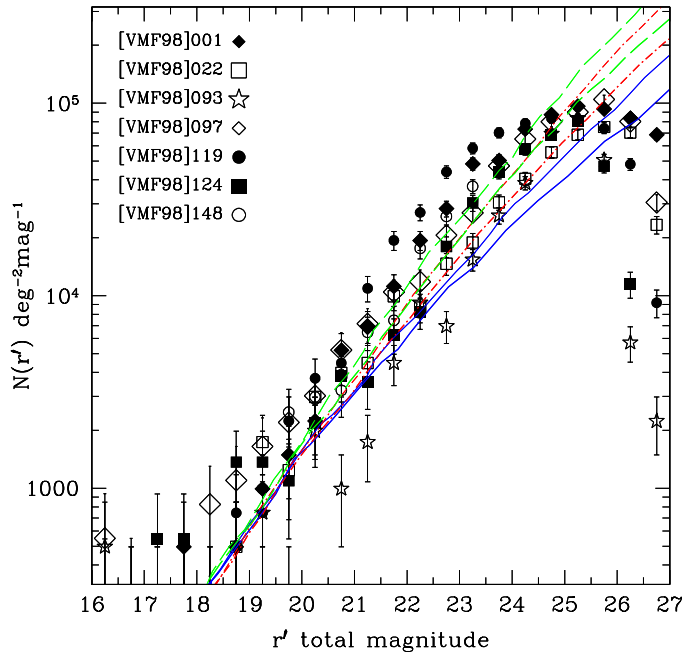}
 \caption{Number counts in the r$^\prime$ band for the galaxy clusters. 
 Models from Nagashima et al. (2002) are included. Thick (thin) lines represent
 those models with (without) selection effects.  The three models: SC, OC and LC are
 exhibited with different line types (see the text for details).}
 \label{ncount}
 \end{figure}

 Figure~\ref{ncount} shows
  r$^{\prime}$ number counts on the logarithmic scale for all the galaxies in
 the GMOS field of view for the seven galaxy clusters, with error bars 
being estimated
 using Poisson uncertainties.  Models from Nagashima et al. (2002) are also 
 shown, with those with selection effects being represented by  
 thick lines, while models without selection effects are shown by thin lines.
A large scatter can be observed in the number counts at brighter magnitudes and 
a strong decay beyond 25 mag. Also, an excess at fainter magnitudes 
may be seen associated with the blue faint 
 excess found by Bruzual and Kron (1980) and King and Ellis (1985). 

In general, the observed number counts were slightly higher
than those from the models, as expected due to the presence of clusters in the
fields.  
The number counts at a fixed $r^{\prime}$ = 22 mag 
range between 5$\times$10$^3$ and 2$\times$10$^4$ (clusters [VMF98]093 
and [VMF98]119,
respectively), consistent with their X-ray luminosities (see Table\ref{table1}).

Linear regressions were also obtained of the number counts between 
 19 $<$ r$^{\prime}$ $<$ 24 with a slope in the range 0.3 to 0.43, 
consistent
 with the typical value of about 0.37 found by Metcalfe et al. (2001) in the
 regime of 20 $<$ R $<$ 26 mag. We noticed that the different passbands
 and adopted techniques, along with the assumptions and completeness 
corrections made it 
 difficult to compare these results with other authors.

 \subsection{The Colour-Magnitude Relation}

 Figure~\ref{cmd} shows the (g - r)$^{\prime}$ CMRs for the 
clusters [VMF98]097 and [VMF98]124, and in Figure~\ref{cmd1} are the 
(r - i)$^{\prime}$
CMRs for [VMF98]001, [VMF98]022, [VMF98]093, [VMF98]119 and [VMF98]148. 
Small black dots 
 represent objects classified as 
 galaxies according to our photometric criteria (Paper I), implying that had
a semi-minor to semi-major axis ratio higher than 0.9, an SExtractor CLASS\_STAR
 smaller than 0.8 and a half-light radius bigger than 5 arcsec. 
 Big black dots correspond to galaxies with photometric redshifts, with
open squares representing members of the three
 galaxy clusters after applying the spectroscopic redshift criteria.  This
resulted in 26 
confirmed  galaxy members for [VMF98]022, 37 for [VMF98]097 and 12 for 
[VMF98]124 (Paper I).  Open triangles represent cluster members 
defined using photometric redshift criteria, with 22, 14, 20 and 16 members 
being found for the 
clusters [VMF98]001, [VMF98]093, [VMF98]119 and [VMF98]148, respectively. 
Vertical dashed lines 
 correspond to our limiting magnitudes with 90\% and 50\% completeness 
 levels, as defined in Paper I.  The linear RCS regressions within 
uncertainties, as
described below, are also shown in the Figures.  

 The colours (g - r)$^{\prime}$
 in the redshift range of 0.1 $<$ z $<$ 0.5 and (r - i)$^{\prime}$ for z $>$ 0.5
covered the 4000 \AA~break due to the 
 passband width.  Moreover, this allowed us to 
 separate red early-type galaxies from star forming objects at lower 
redshifts (0.25 to 0.35).  
In this way, we were able to distinguish the red sequence and star 
forming galaxies (Bruzual 1983; Dressler \& 
 Shectman 1987).  K-corrections were not applied to our magnitudes as we had 
 photometric 
 information in only two passbands and therefore corrections would have been
approximate. 
 Also, using tabular values, as for instance those from Frei \& Gunn (1994)
 would only give rough estimates due to uncertainties in our morphological 
 estimates.

 \subsubsection {Red cluster sequence}

\begin{figure*}
 \includegraphics[width=55mm,height=55mm]{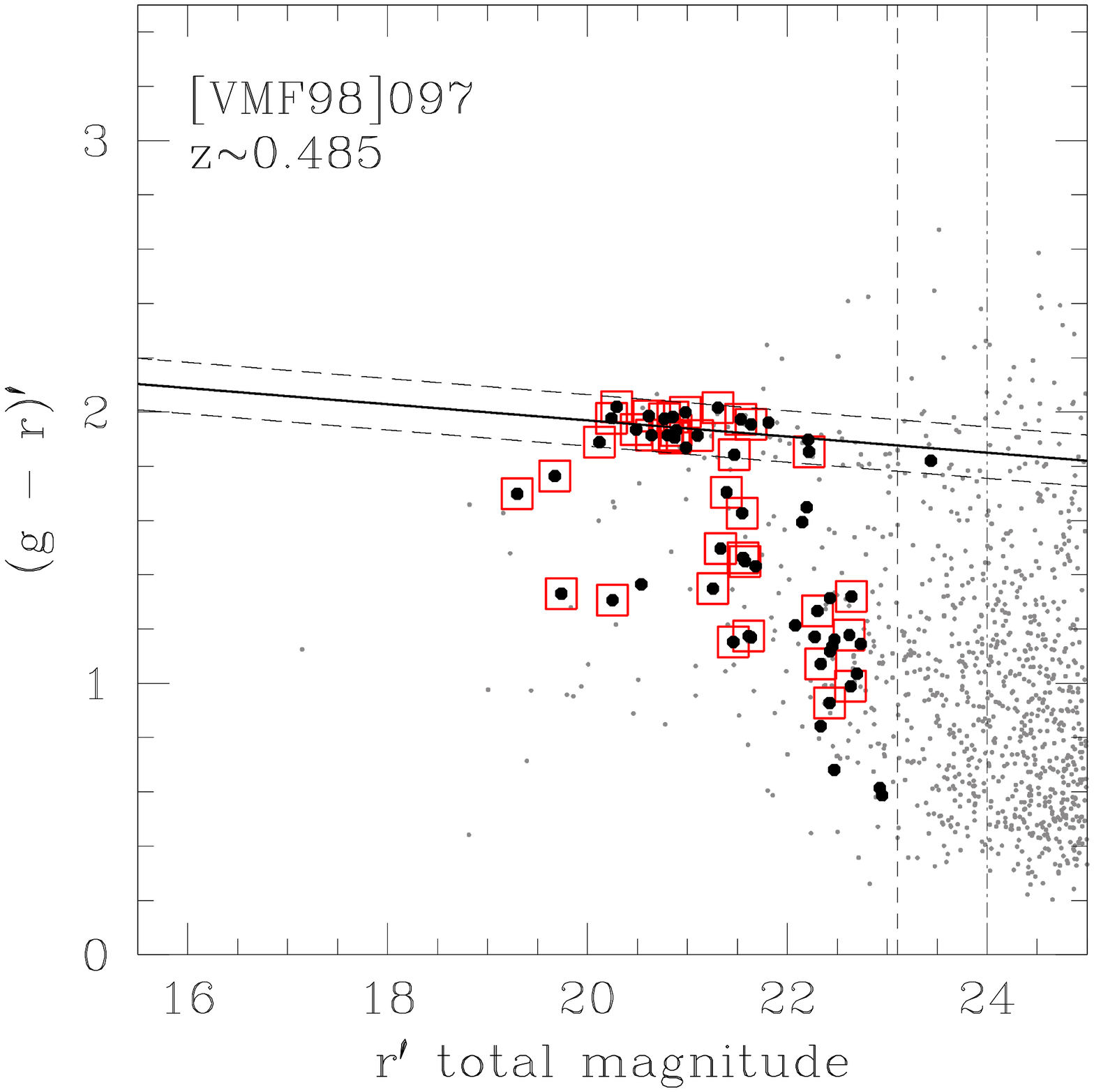}
 \includegraphics[width=55mm,height=55mm]{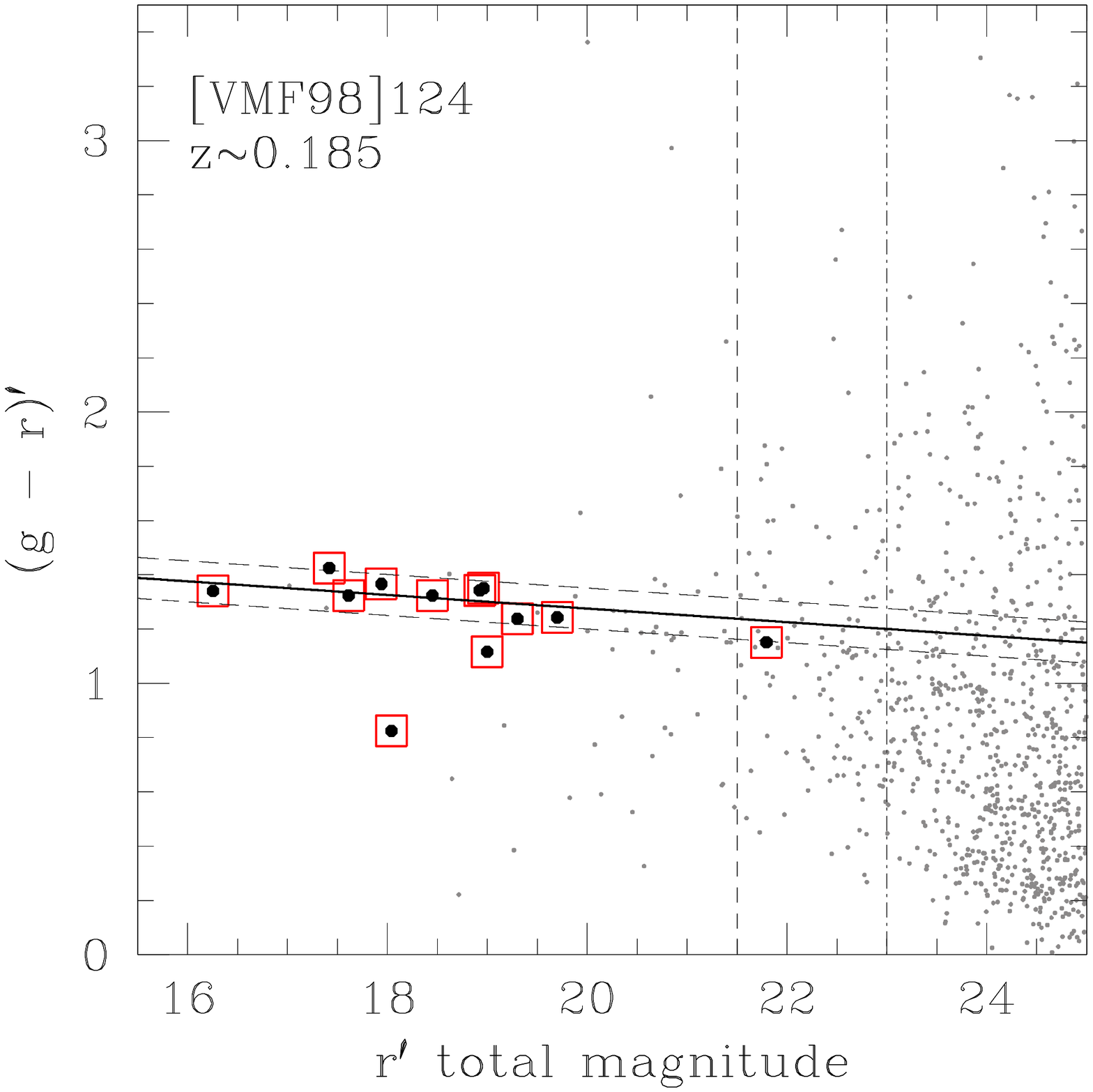}
 \caption{(g - r)$^{\prime}$ Colour - Magnitude Relations in the 
 neighbourhoods of the galaxy clusters.  Small grey dots 
 represent objects classified as 
 galaxies. Big black dots correspond to galaxies with photometric redshifts. 
Open symbols are those confirmed cluster members
 with spectroscopic data (squares) or photometric redshifts (triangles).
Vertical dashed lines 
 correspond to limiting magnitudes with 90\% and 50\% completeness 
 levels. The linear RCS regressions (solid line) within uncertainties are 
also shown.}
 \label{cmd}
 \end{figure*}

\begin{figure*}
 \includegraphics[width=55mm,height=55mm]{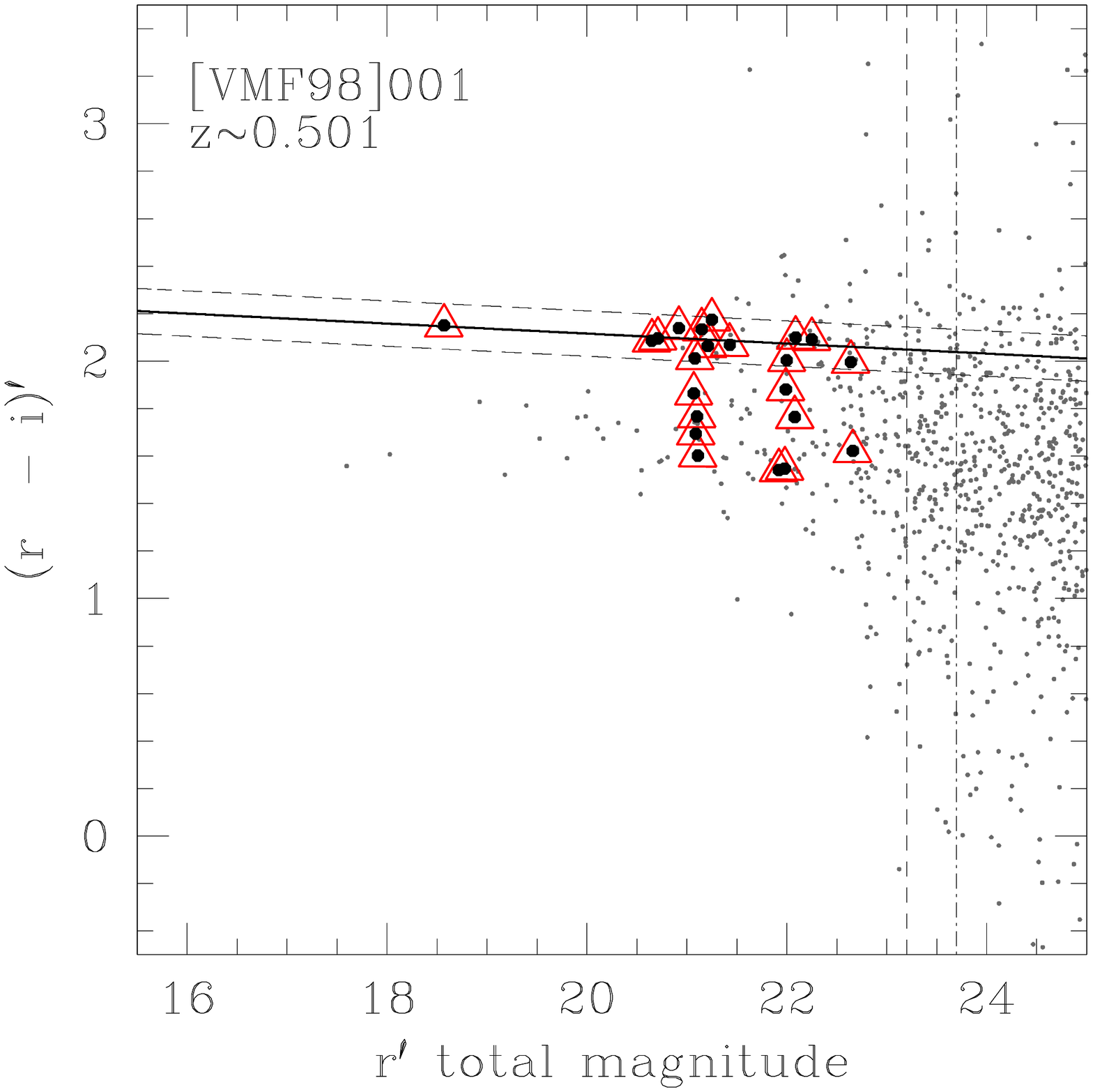}
 \includegraphics[width=55mm,height=55mm]{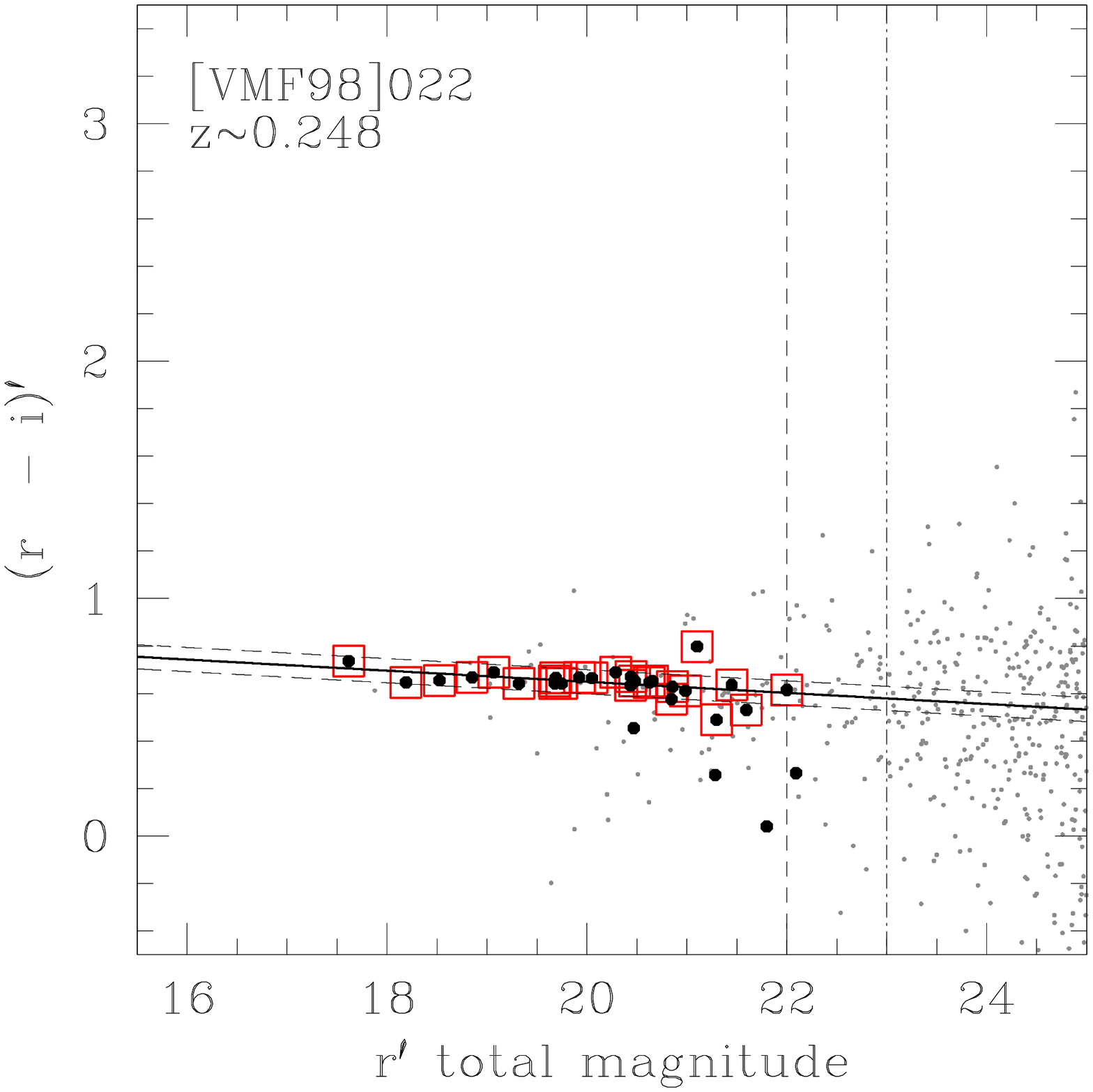}
 \includegraphics[width=55mm,height=55mm]{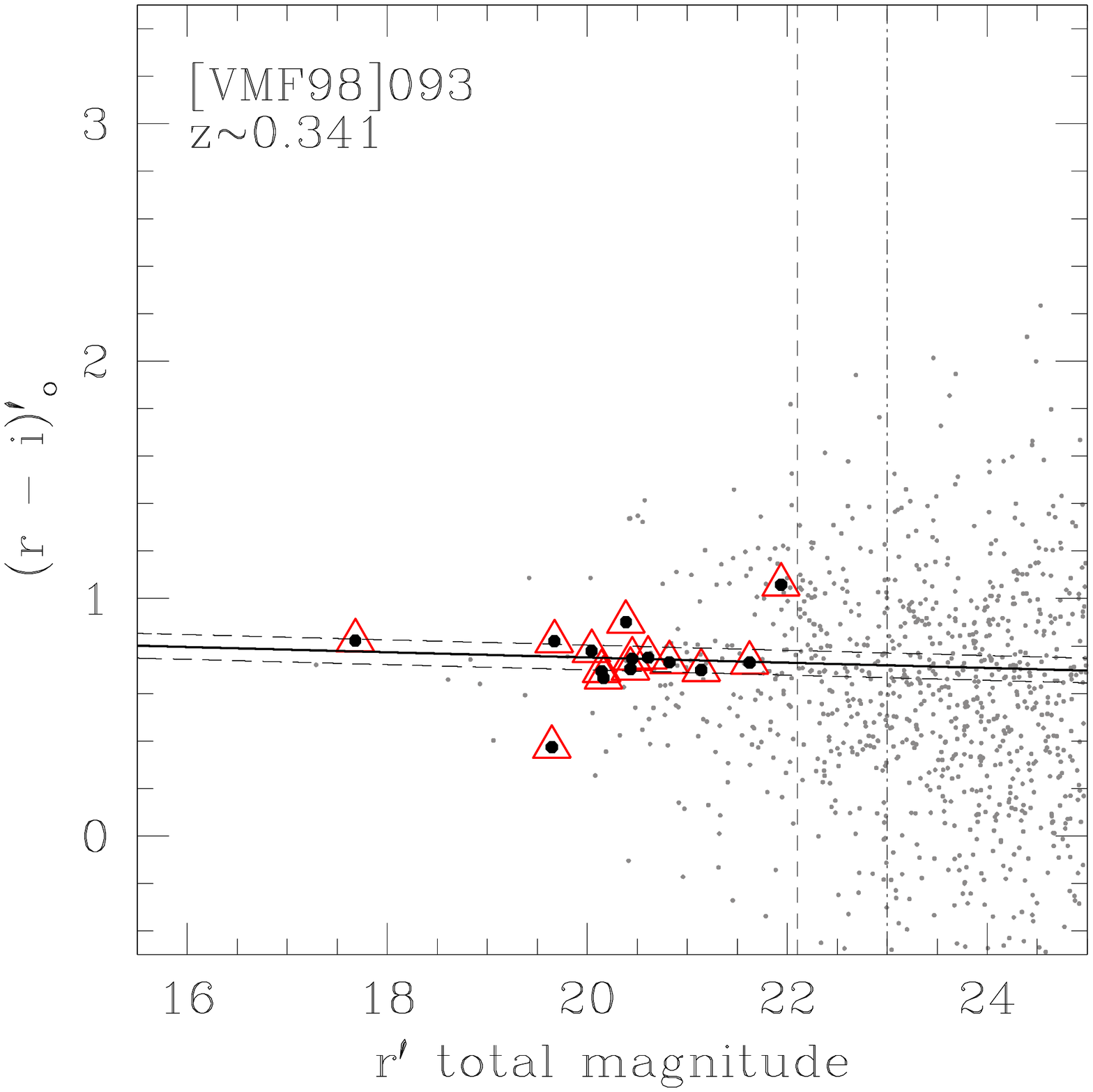}
 \includegraphics[width=55mm,height=55mm]{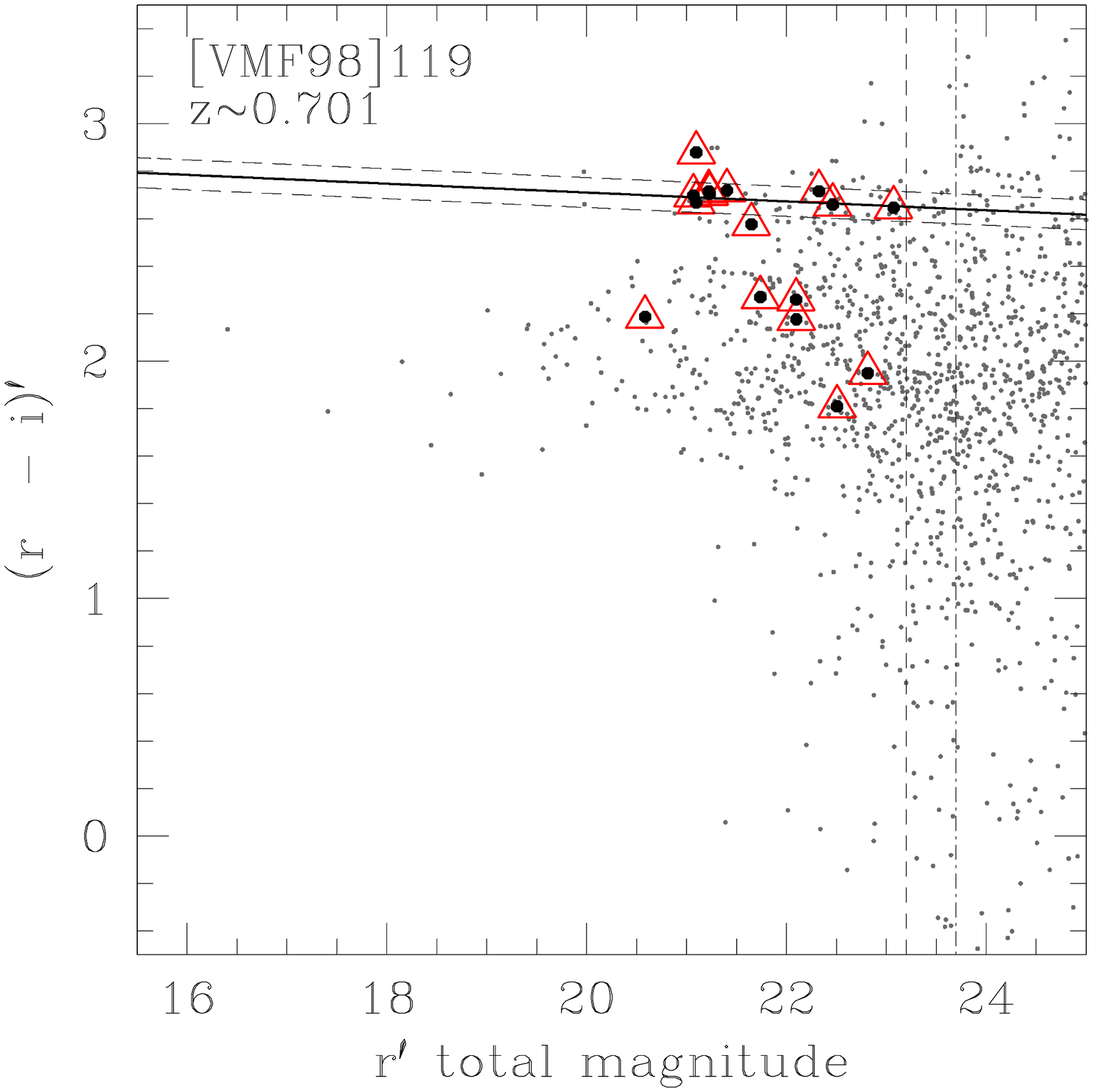}
 \includegraphics[width=55mm,height=55mm]{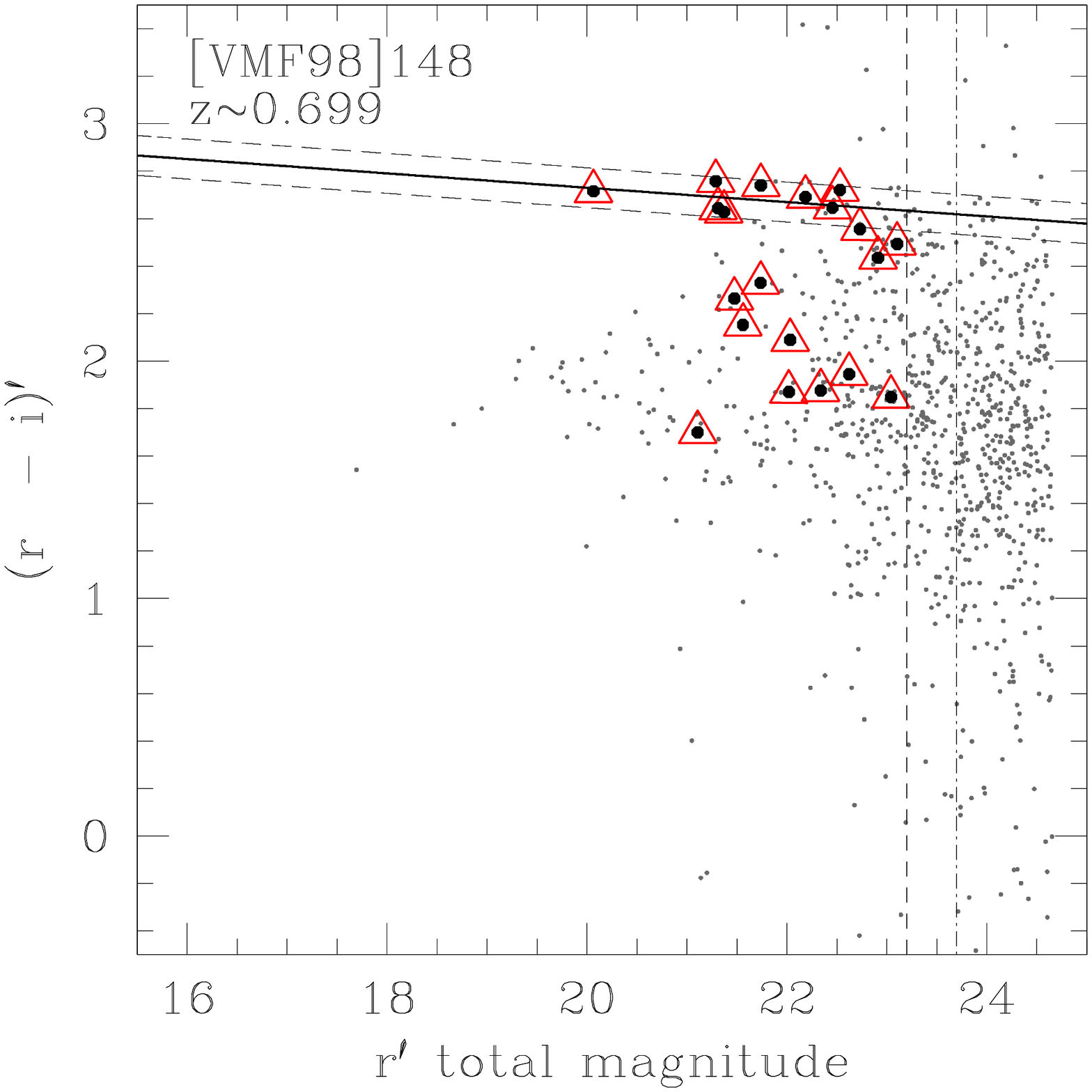}
 \caption{(r - i)$^{\prime}$ Colour - Magnitude Relations in the 
 neighbourhoods of the galaxy clusters.  Symbols are explained in previous 
Figures.
 }
 \label{cmd1}
 \end{figure*}

On inspection of the CMRs of Figures~\ref{cmd} and \ref{cmd1}, a variety of red
cluster sequences can be observed. For low redshift clusters, the 
presence of a well defined  RCS with an 
 extension of 4 to 6 magnitudes may be seen.  This trend 
 is the main characteristic of virialized 
 systems with a dominant elliptical 
 galaxy, which has low signatures of star formation.  However, in cluster 
 [VMF98]124 a range of about two magnitudes 
without cluster members can be observed, with
 comparable RCS results also being found by 
 Gladders et al. (1998) in their sample of lower redshift clusters and by 
 Stott et al. (2009) and Gladders et al. (2005). Steeper slopes were reported
by Wake et al. (2005) for the RCS in poor galaxy clusters at lower redshifts,
with the dispersion attributed to uncertainties in the definition of galaxy 
members.

 Mei et al. (2006a, 2006b) studied massive galaxy clusters at higher 
 redshifts (z $>$ 1) and focused on the RCS.  
 Their reported slopes ranging -0.02 to -0.03 with the early-type red 
 sequence being well defined out to redshifts of about 1.3 (Mei et 
 al. 2009).  For the higher redshift clusters we found an RCS with 
an extension of about 2 to 4 
magnitudes, with these clusters presenting a different behaviour than the lower 
redshift clusters, revealing a 
mix of early-type galaxies together with more blue galaxies. 
This result is in agreement with that reported by 
 De Lucia et al. (2007) in a study on massive clusters at z $>$ 0.8 and that
of a galaxy cluster at z $>$ 1 reported by Lerchster et al. (2011). 
 At these redshifts, however, it is difficult to compare results from 
different studies, because most of these were based on massive galaxy clusters.

As Gladders et al. (1998) noted, a lot of care has to be taken in the 
selection of the RCS galaxies, with, the total extension of the RCS and its
 scatter having been discussed by several authors for different 
 types of clusters. For instance, Terlevich et al. (2001) and Secker, Harris 
 \& Plummer (1997) defined an 8 
 magnitude RCS for the nearby clusters of Virgo and Coma.  At redshifts of 
 $\sim$0.15, 
  Carrasco et al. (2006) determined the galaxy cluster RCS to be of 6 
 magnitudes and even reaching 10 
 magnitudes with a set of 4 groups and fossils.  For low mass clusters, Balogh 
 et al. 
 (2009) reported a 6 magnitude RCS using a spectroscopic sample of 10 
groups between z$\sim$0.2 and 0.3. In addition, at higher
 redshifts, De Lucia et al. 
 (2006) identified an RCS extension of 6 magnitudes for clusters at 
 z$\sim$0.5, and of 4 magnitudes for those at z$\sim$0.8, using the 
 EDISCS sample.  Based on these previous results, we chose to define the RCS as
 those cluster members which were 3 magnitudes fainter than the second 
brightest galaxy in
 the cluster, avoiding the BCG.  This limit 
 corresponded to approximately a 90\% magnitude completeness level, which is 
 represented by vertical short dashed lines in the panels of the Figures.
 For those clusters with a less clear RCS (see Figures~\ref{cmd} and 
\ref{cmd1}), 
 we selected early-type objects using 
 only the redder galaxies
 at the cluster core which had a projected radius of about 0.3 Mpc 
(following Sarazin 1986),
 concentration index higher than 2.5, and ellipticities lower than 0.4 or 
0.7 $<$ b/a $<$ 0.9.
In this way, we studied the central parts of the clusters dominated by 
 red-spheroidal galaxies, thus reducing foreground/background contamination.

 Several authors have studied the performance of different techniques to fit 
 the RCS, to provide estimates of the slope and zeropoint (Press et al. 
1992, van Dokkum et
  al. 1998, L\'opez-Cruz et al. 2004, Stott et al. 2009, Hao et al. 2010). 
 Gladders et al. (1998) argue that linear regression plus an iterated 
 3 $\sigma$-clipping provides the most stable results.  We followed
  this technique, but applied a 1 $\sigma$-clipping iteration to create 
 a more conservative RCS sample. Then, using a least-square regression, 
 the RCS slope and zeropoint were obtained, with errors being estimated 
using 10.000 bootstrap re-sampling.
 Table~\ref{table4} 
 shows the RCS number of galaxies
 and the slope and zeropoint with their 
 corresponding uncertainties for each cluster. The derived RCS slopes had 
a mean of approximately 0.02 with a spread of about 0.01.

 Taking into account the linear fit regressions, for each cluster we defined
the galaxy samples with colours over
 and below (within 1$\sigma$) the RCS sample.  Hereafter, the
 Blue sample is defined as those galaxies with colours below the RCS, and 
will be used for further analysis in the following sections.

 \subsubsection {Colours}

\begin{figure*}
 \centering 
 \includegraphics[width=60mm,height=60mm]{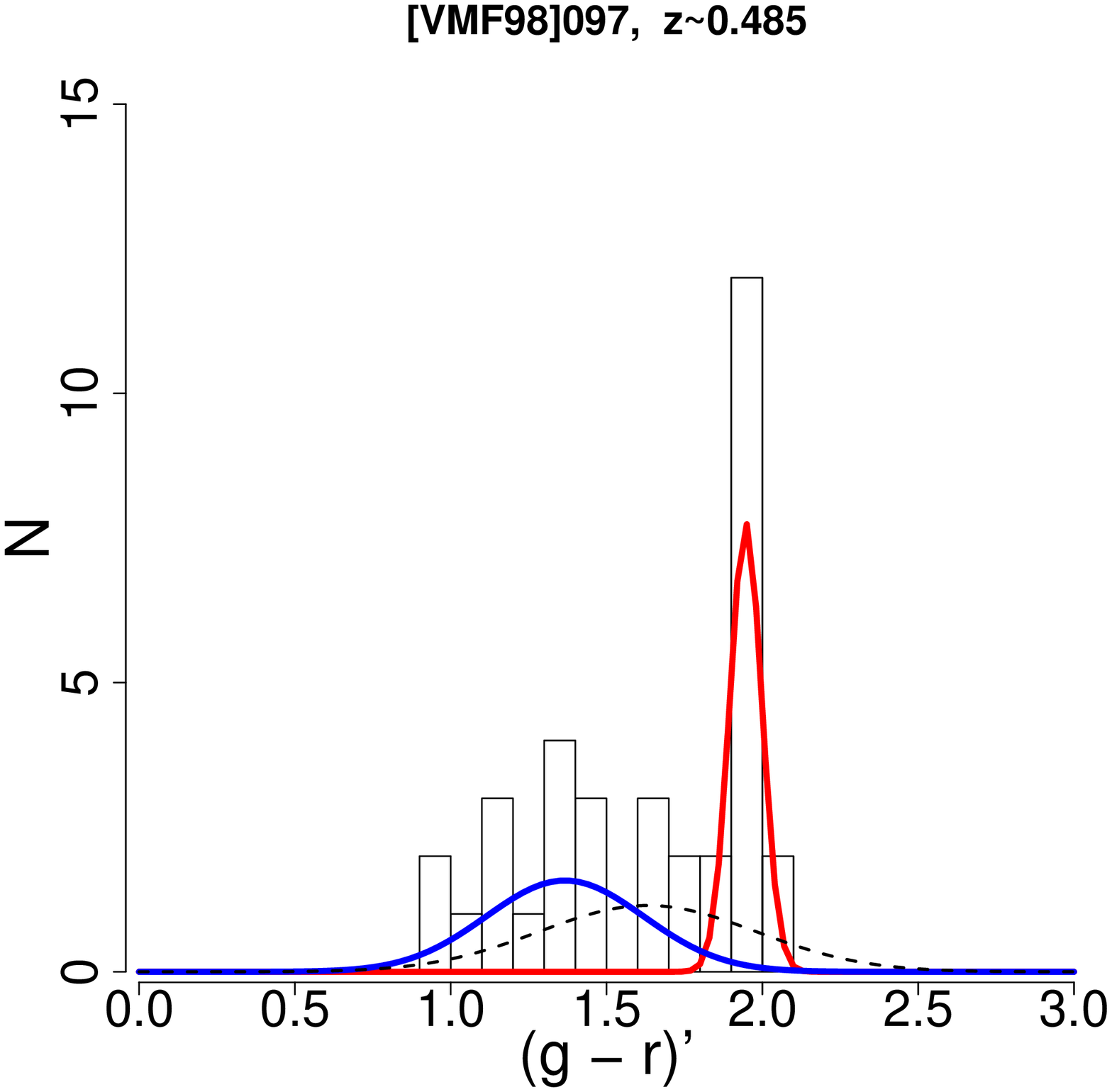}
 \includegraphics[width=60mm,height=60mm]{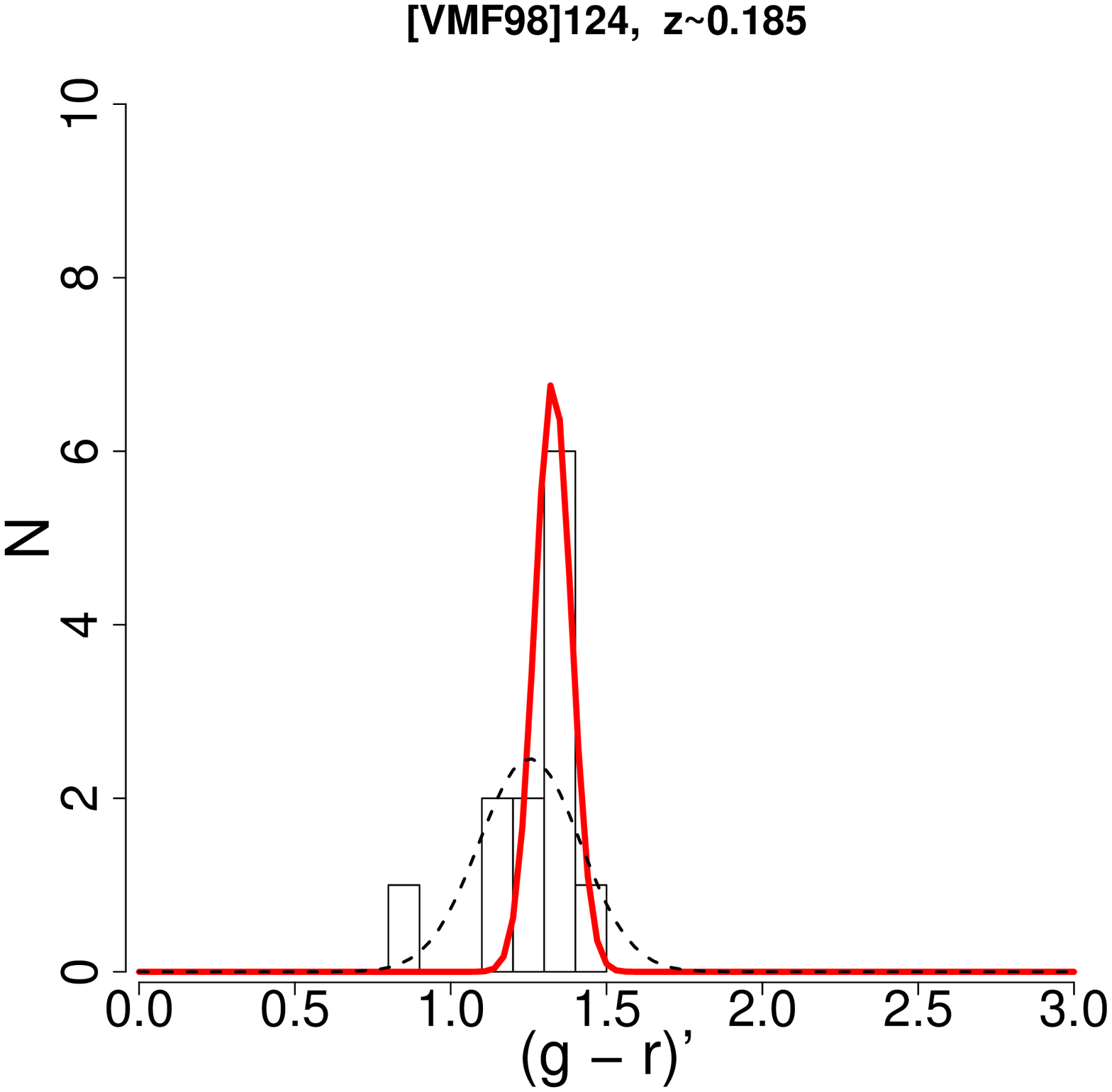}
 \caption{(g - r)$^{\prime}$ colour distribution for the cluster galaxy members.
Single fits are displayed: dashed black distribution represents the total 
member galaxies, in colours are shown the red and blue galaxy samples.}
 \label{color1}
 \end{figure*}

\begin{figure*}
 \centering
 \includegraphics[width=60mm,height=60mm]{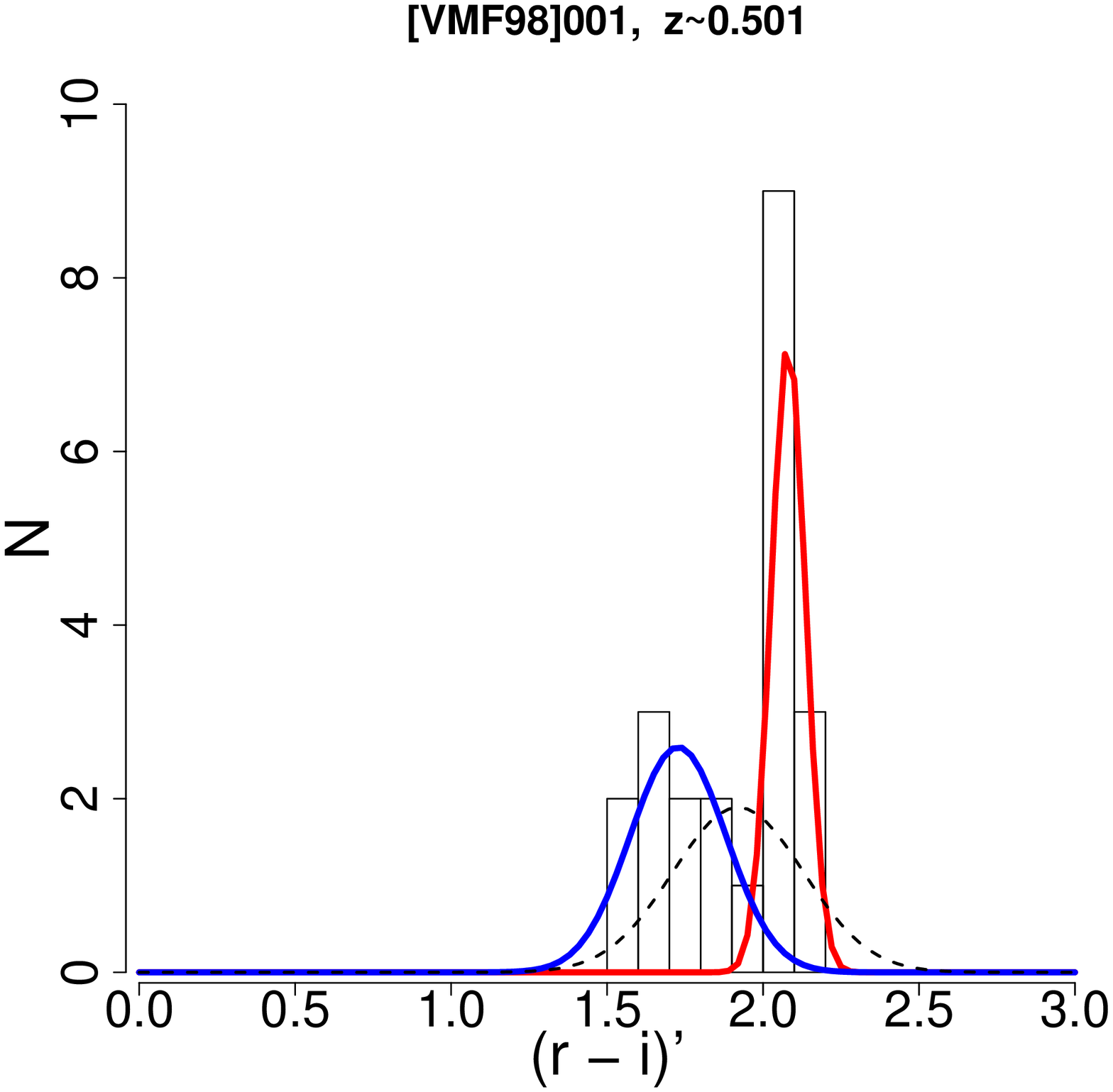}
 \includegraphics[width=60mm,height=60mm]{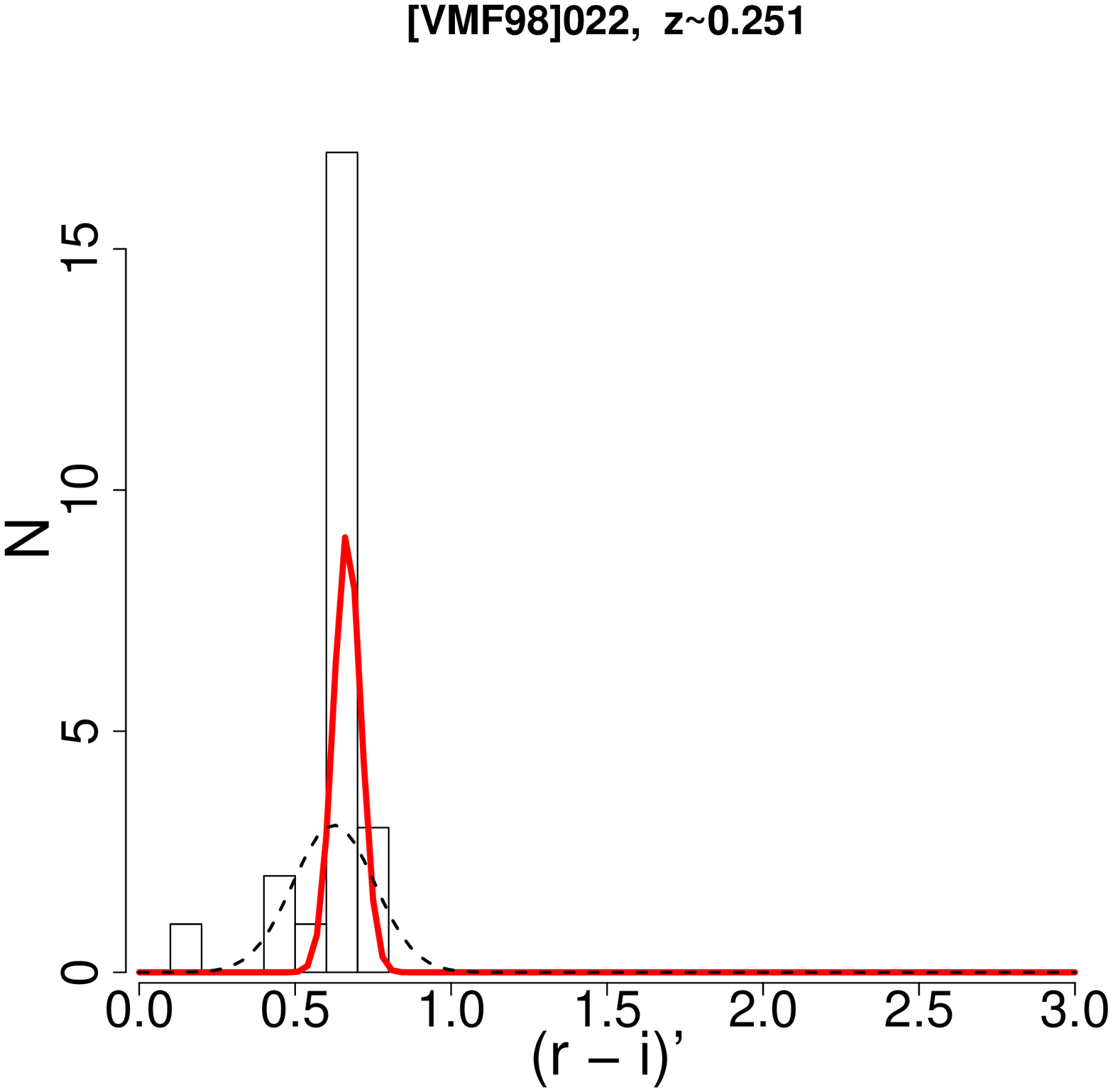}
 \includegraphics[width=60mm,height=60mm]{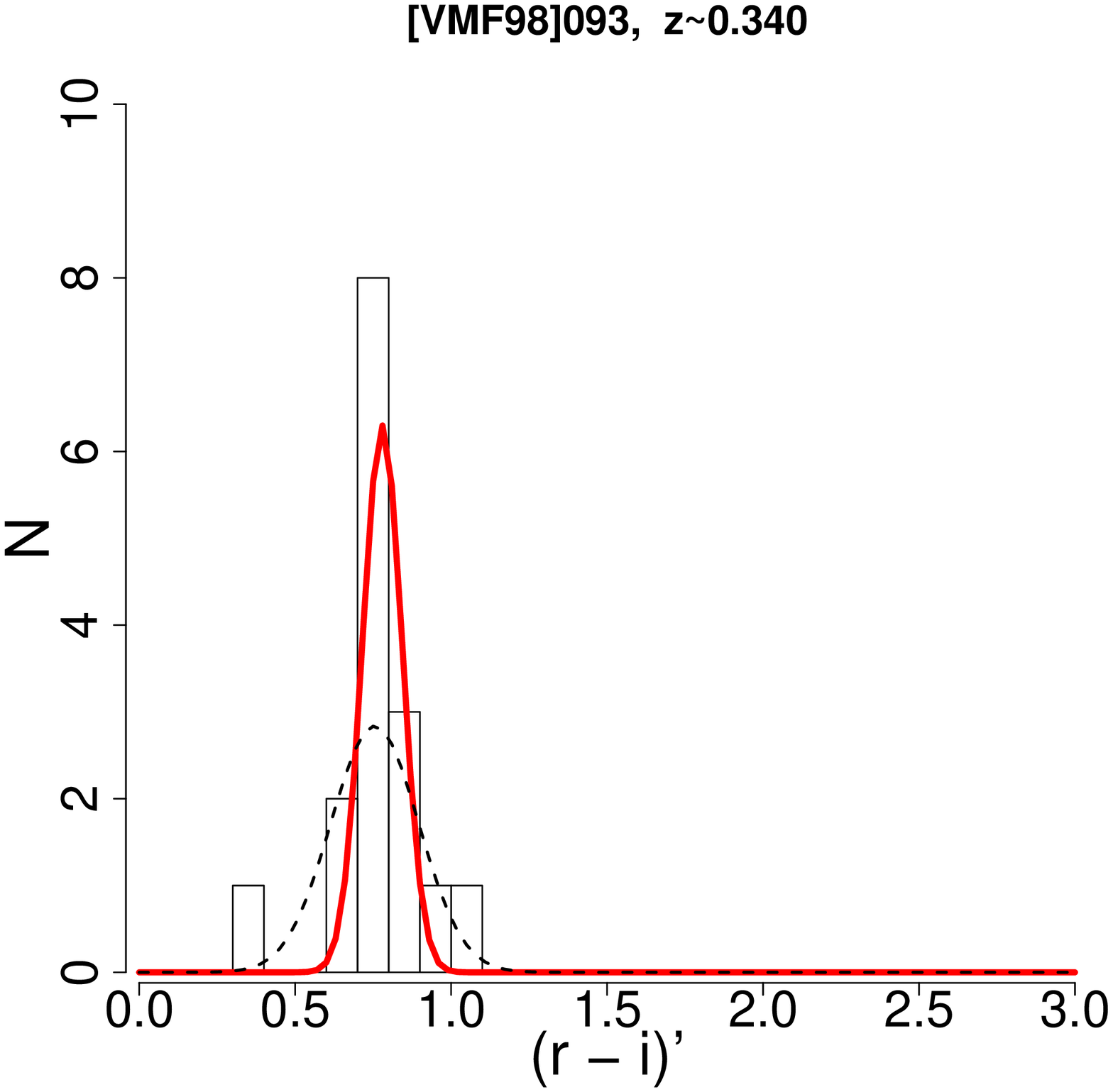}
 \includegraphics[width=60mm,height=60mm]{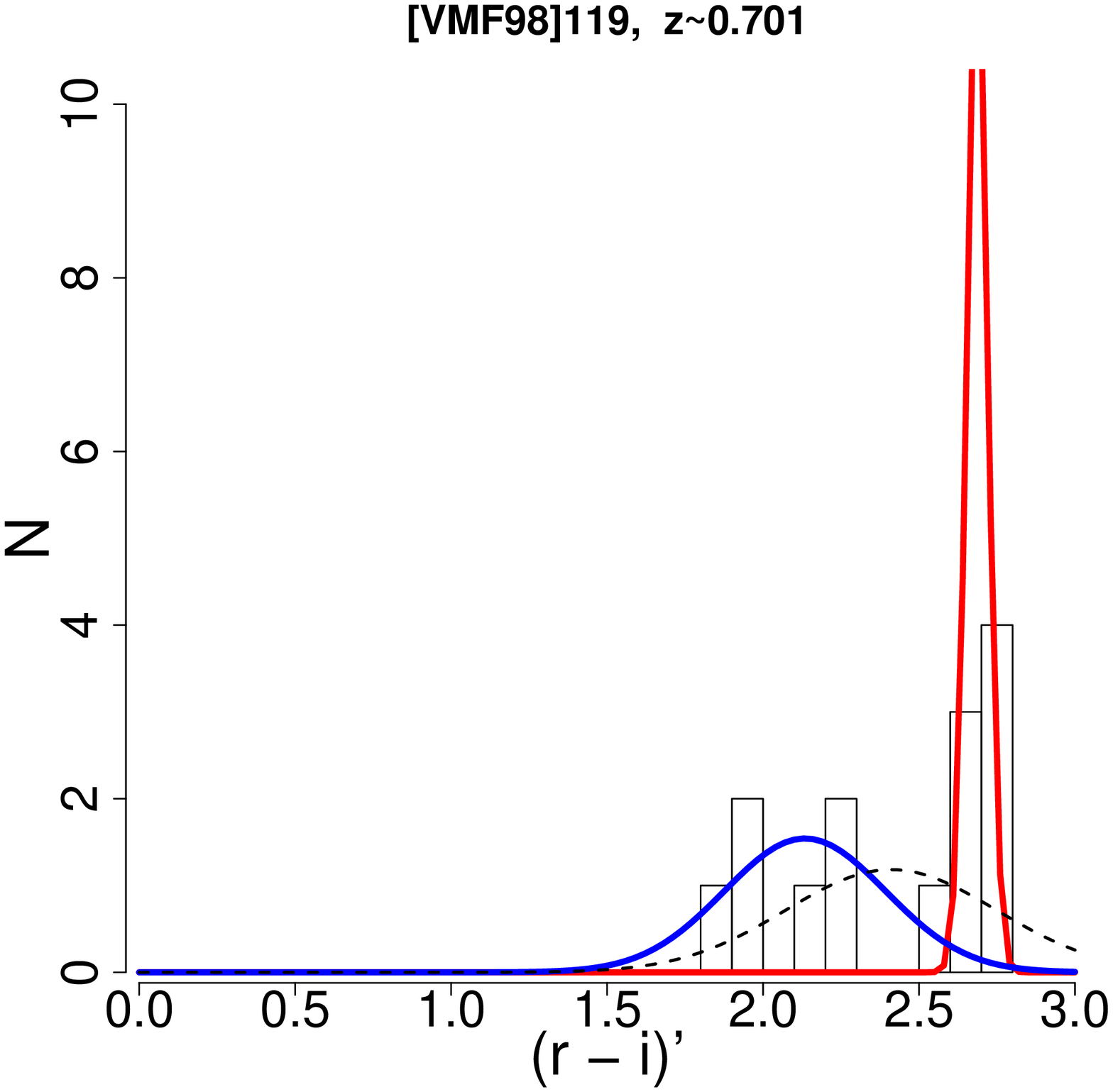}
 \includegraphics[width=60mm,height=60mm]{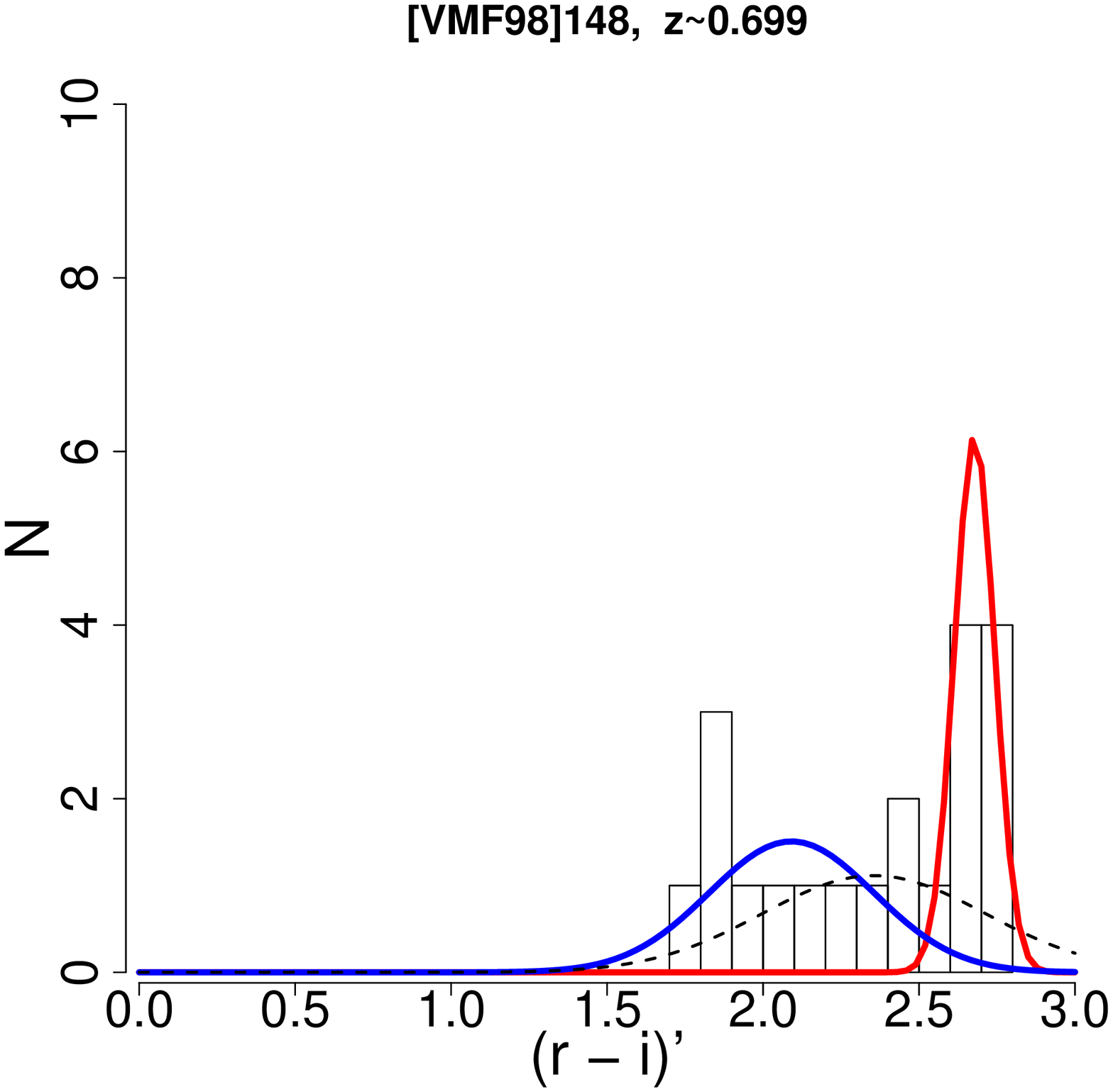}
 \caption{(r - i)$^{\prime}$ colour distribution for the cluster galaxy members.
Single fits are displayed as in previous Figures.}
 \label{color2}
 \end{figure*}

 The colour distribution of the galaxy members was analysed for 
 the seven clusters studied. For each distribution, we 
performed a statistical 
 analysis considering both a single distribution for the whole galaxy sample 
and two Gaussian distributions for the RCS and blue samples.  
Figures~\ref{color1} and ~\ref{color2} show the colour 
distributions for the cluster members with the dashed black curve
corresponding to the single Gaussian distribution, while the blue and red 
curves indicate the two distributions.   
It was found that the single distribution was not a good representation of the
colour distribution, and it was not possible to fit the blue population with
reasonable precision for the three low redshift clusters [VMF98]022; 
[VMF98]093 and [VMF98]124 due to the small number of blue galaxies.  
The observed (r - i)$^{\prime}$ colours in the first two clusters may have been
responsible for this small number in this low redshift range. 
However, for [VMF98]124, the lowest redshift studied cluster, 
the GMOS FOV allowed us to study in detail the 
central parts by avoiding the blue galaxies in the outer parts.

Some galaxies had extreme colour values of around
(g - r)$^{\prime}$ $\sim$ 0.8 in [VMF98]124 and 
(r - i)$^{\prime}$ $\sim$ 0.37 in [VMF98]093, which corresponded to 
approximately 5$\sigma$ of the RCS fit (Figure~\ref{cmd}).   
In general, the blue excess was a good indicator of 
 strong star formation in galaxies, with this effect possibly being related 
to cluster 
 dynamics.  McIntosh et al. (2004) studied extreme colour galaxies and 
reported that they are spatially, kinematically, and morphologically 
distinct from 
 red cluster galaxies.  Despite the colour of these galaxies, they were
 spectroscopically confirmed as cluster members, and are therefore
 genuine members of the original samples. 

 Table~\ref{table5} shows the statistical results for the red and blue 
populations: mean values, standard deviation and $\chi$$^2$ of the colour
distribution of galaxy members.  Despite the differences in the observed 
filters, there was an absence of blue galaxies at lower redshifts, with
a red population occurring with a peak at (g - r)$^{\prime}$ $\sim$ 1.33 for 
cluster 
 [VMF98]124 and (r - i)$^{\prime}$ $\sim$ 0.64 and 0.76 for clusters 
 [VMF]022 and [VMF98]093, respectively.  At higher redshifts, 
there were
two clear colour populations in the four galaxy 
clusters, probably due to a mix of red and blue galaxy 
 populations, thus suggesting a dynamically active cluster in the case of
[VMF98]097 (Carrasco et al. 2007).

 \subsubsection {The Butcher-Oemler effect}

\begin{figure}
 \centering
 \includegraphics[width=80mm,height=80mm]{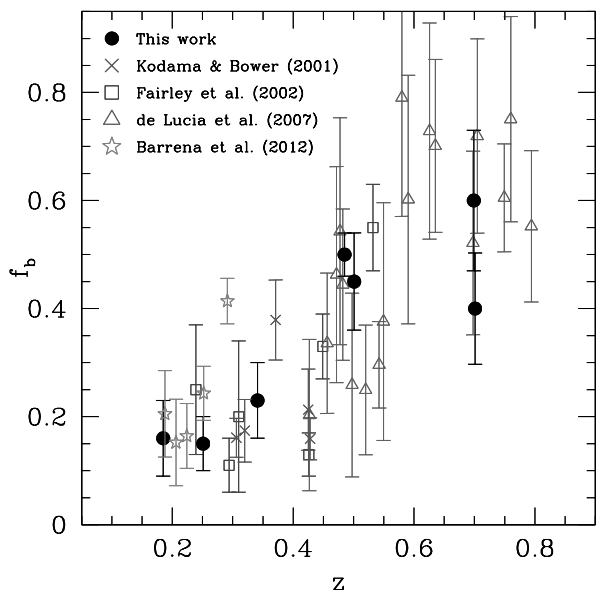}
 \caption{Butcher-Oemler effect for the studied galaxy clusters (filled circles).  
 Blue fraction from galaxy clusters with X-ray luminosity of about 
10$^{44}$ ergs s$^{-1}$ from Kodama \& Bower 
(2001, crosses), Fairley et al. (2002, squares), de Lucia et al. (2007, 
triangles) 
 and Barrena et al. (2012, stars) are also included. }
 \label{fb}
 \end{figure}

 This effect (Butcher \& Oemler, 1978) is related to the fraction 
 of blue galaxies ({\it f$_b$}) in clusters, and its evolution over the last 6 
 Gyrs. We calculated {\it f$_b$} in our galaxy clusters as the number of 
 blue galaxies in the total distribution and compared 
 our results with  Kodama \& Bower (2001), Fairley et al. (2002),
de Lucia et al. (2007) 
 and Barrena et al. (2012)
  in the same redshift range (0.1 $<$ z $<$ 0.5), cluster region (r$_p < $ 
0.75 Mpc) and 
 equivalent colours.
 Figure~\ref{fb} shows the blue fraction as a function of redshift for our 
cluster sample (filled circles) along with those  
from literature. For both our data and those of the literature, 
a consistent
 increase can be observed in the blue fraction of galaxies from lower redshift
clusters compared to those at higher redshifts.
These values were related with a 
 97\% confidence using the Kendall $\tau$ rank correlation coefficient.
This larger
 fraction of blue galaxies in higher redshift clusters can 
also seen in Figures~\ref{cmd} and ~\ref{cmd1}.  The larger uncertainties in 
higher redshift clusters could be due to 
 sample selection or a larger mass spread of the clusters, among other
factors.

 \subsection{Galaxy Number Density Profiles}

\begin{figure*} 
 \includegraphics[width=55mm,height=55mm]{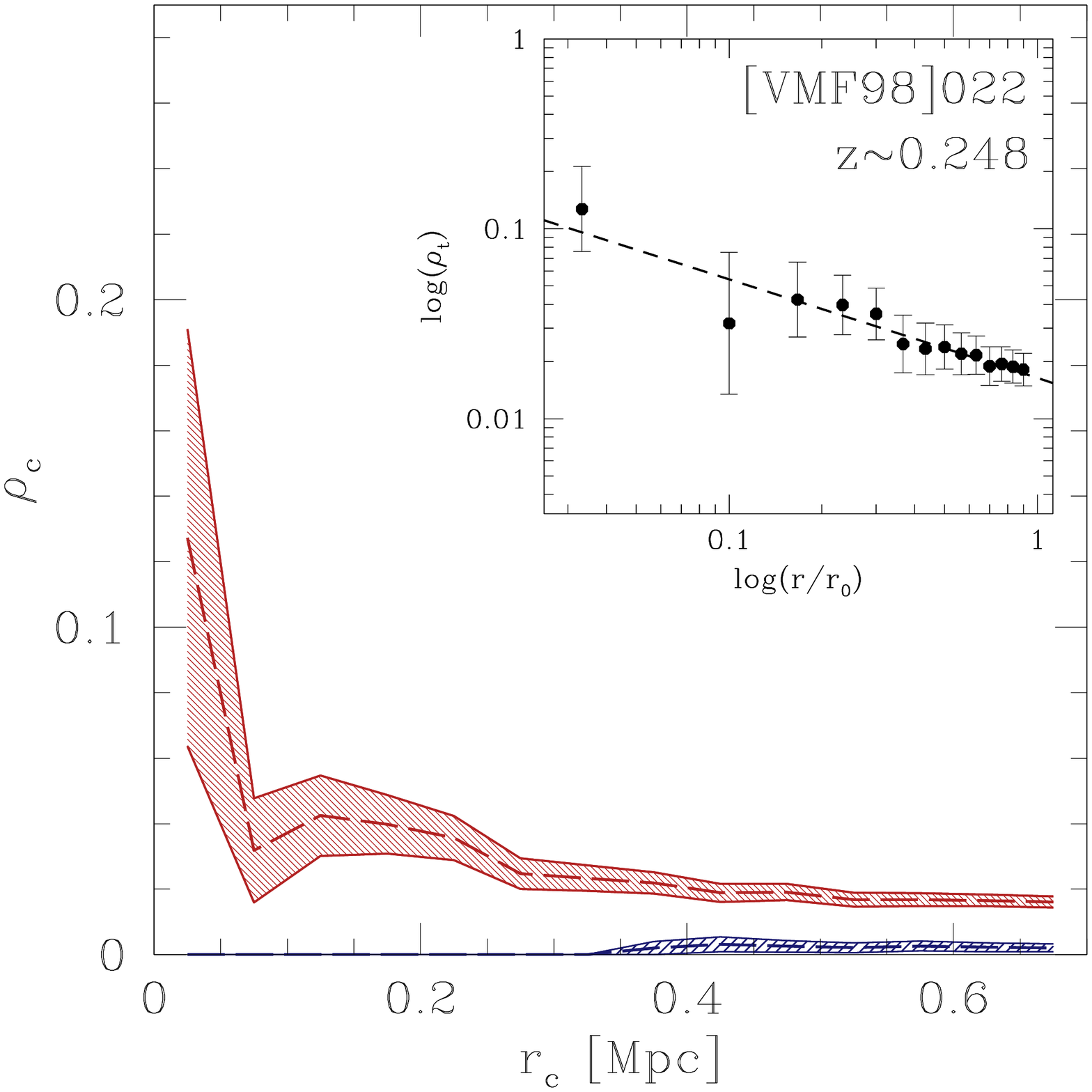}
 \includegraphics[width=55mm,height=55mm]{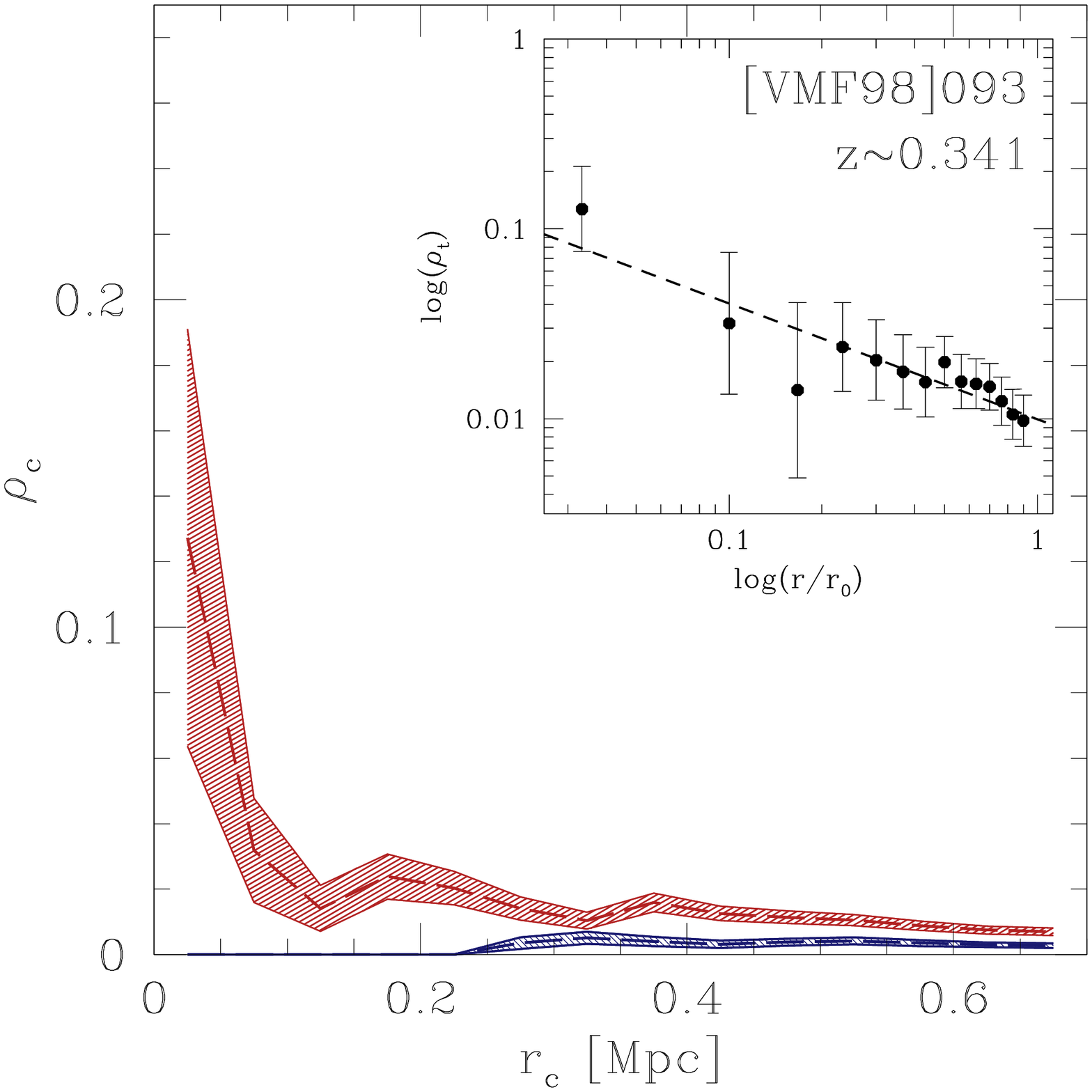}
 \includegraphics[width=55mm,height=55mm]{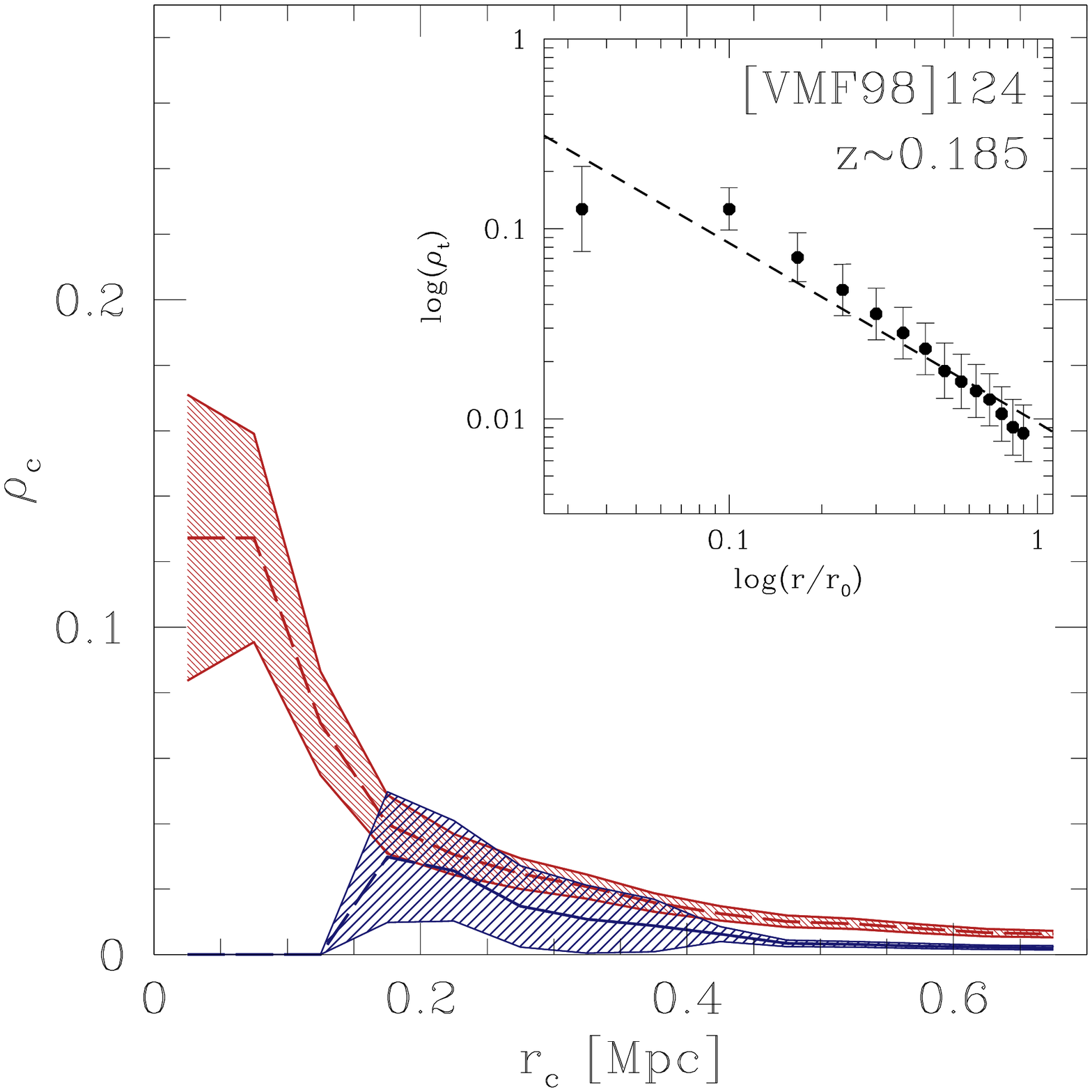}
 \caption{Galaxy number Density Profiles for the low redshift galaxy clusters 
  for the RCS (red) and Blue (blue) samples. The total radial 
galaxy number density profiles are shown
in the small panel (see text for details).}
\label{densityprof1}
 \end{figure*}

\begin{figure*} 
 \includegraphics[width=55mm,height=55mm]{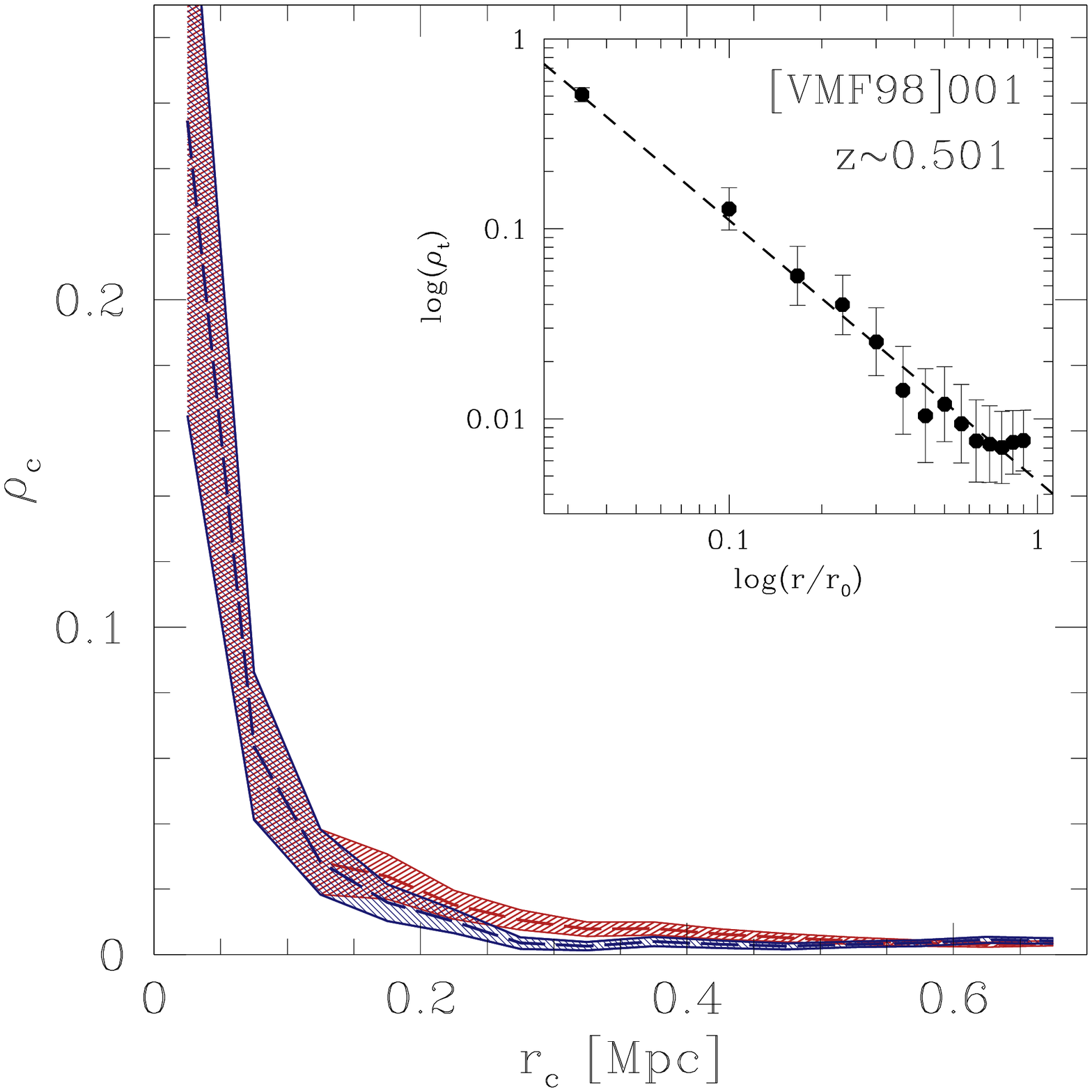}
 \includegraphics[width=55mm,height=55mm]{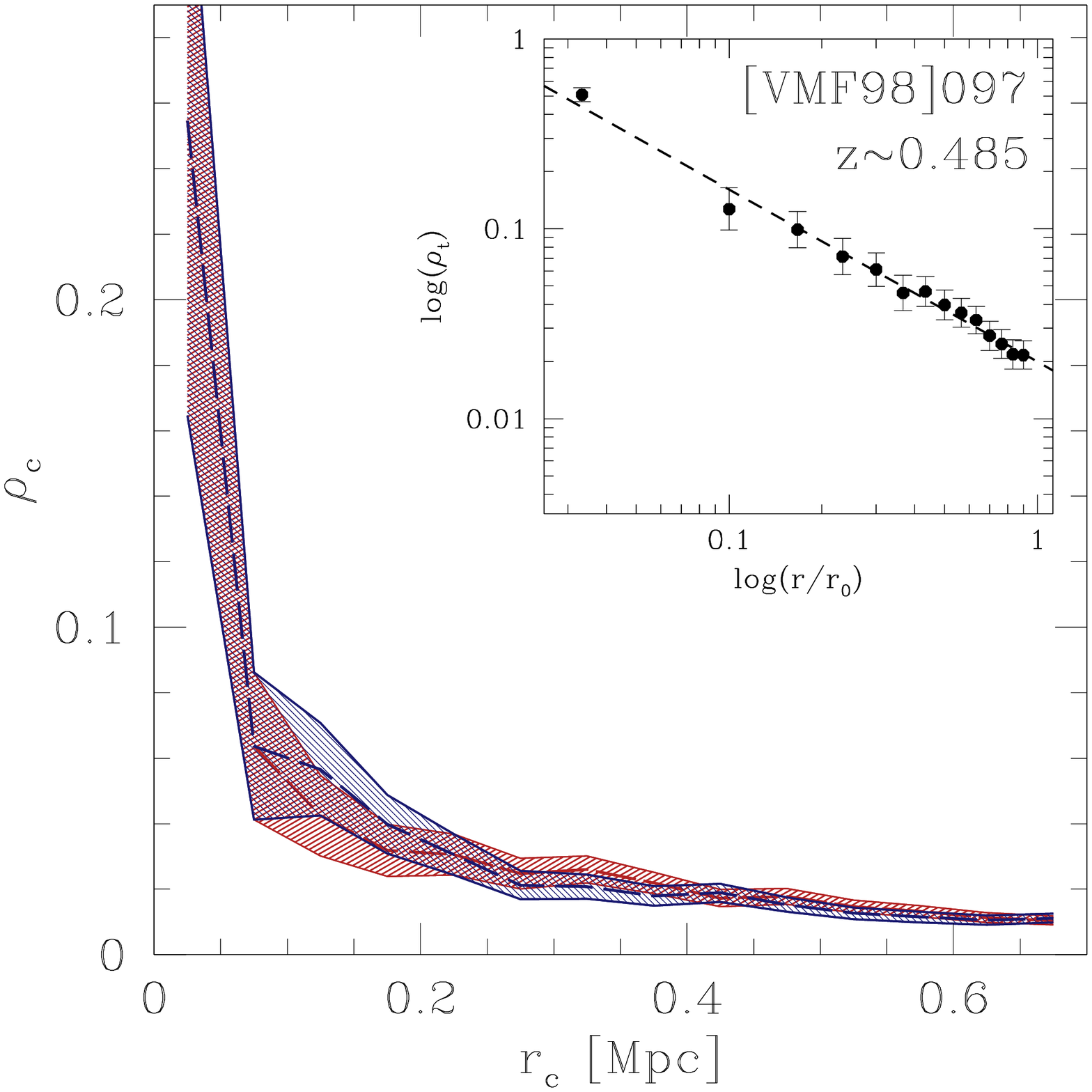}
 \includegraphics[width=55mm,height=55mm]{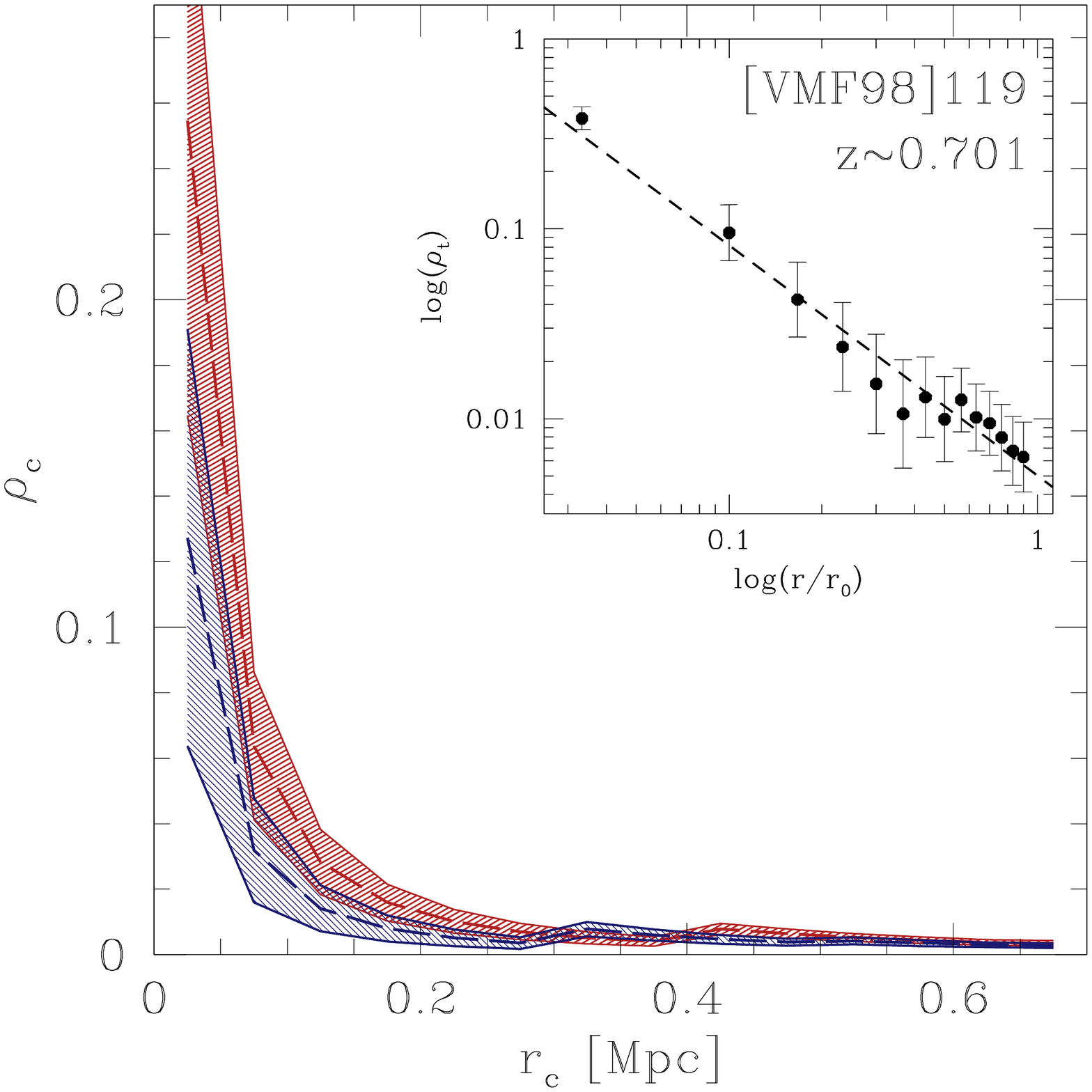}
 \includegraphics[width=55mm,height=55mm]{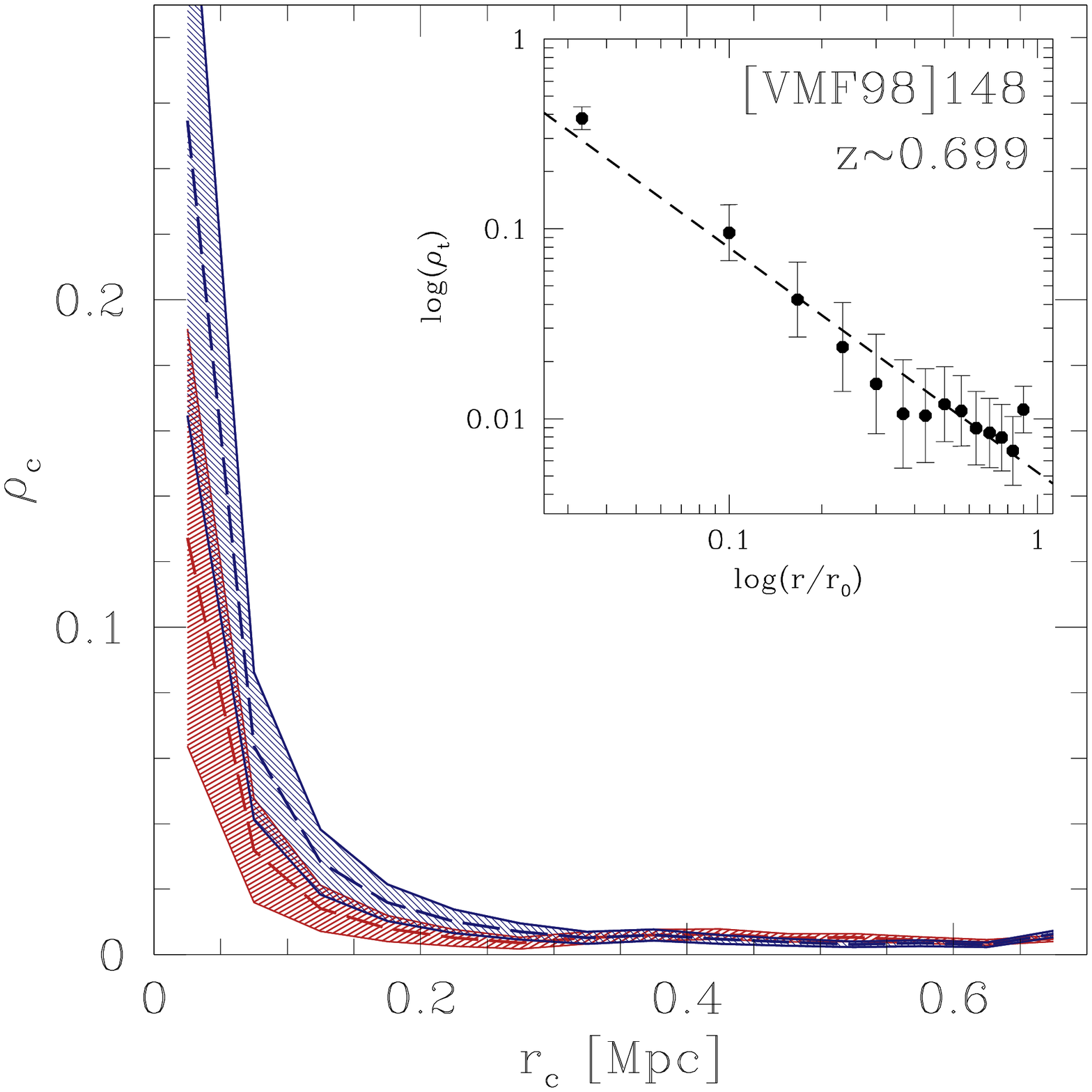}
 \caption{Galaxy number Density Profiles for the high redshift galaxy clusters 
 for the RCS and Blue samples. Panels and colours are as in previous Figures.}
\label{densityprof2}
 \end{figure*}

 The galaxy number density profile was determined
within a total area of radius $\sim$ 1.2 Mpc for our cluster sample.
 For the three lower 
 redshift clusters the centre was defined by the BCG whereas for the higher
redshift clusters it was defined as the peak of the X-ray emission.
 The number density profiles were obtained for the two galaxy samples defined
 in the previous section: the RCS and the Blue samples.  We also calculated the 
 total radial galaxy number density profile for all member galaxies and
we fitted power laws ($\rho (r) = A r^{-\alpha}$) to the radial density 
profiles,
 obtaining $\alpha$ values from -0.65 for [VMF98]022 to -1.37 for [VMF98]001.
Table~\ref{table6} shows these values and the rms for the seven galaxy clusters,
which are in the range reported 
by other authors (typical values from -0.7 to -1.4, Hansen et al. 2005).
Figures~\ref{densityprof1} and \ref{densityprof2} show the galaxy number density ($\rho_c$) profile as 
 a function of the distance to the cluster centre (r$_c$) for low and high 
redshift clusters, respectively. The RCS (in red) and 
 Blue samples are shown within statistical Poisson errors and are 
represented by the 
 shaded areas.  In addition, the total radial number density profile in 
logarithmic scale
 together with the power law fit are included in the inner small panel. 
 A central concentration for 
 the RCS sample with an absence of blue galaxies can be observed for the 
three low redshift clusters.   About 70\% of the red galaxies were located in 
the cluster core (r$_c < $ 0.2 Mpc).  However, a different behaviour was 
noted for the 
high redshift galaxy clusters, where
 a remarkable increase in the number of blue galaxies was
 present, mainly at the centre. This could also be seen as an excess of blue 
 galaxies in the CMDs (Figures~\ref{cmd} and \ref{cmd1}) related to the 
Butcher-Oemler effect (Figure~\ref{fb}).

 \subsection{Galaxy projected distribution} 

\begin{figure*}

 \includegraphics[width=80mm,height=80mm]{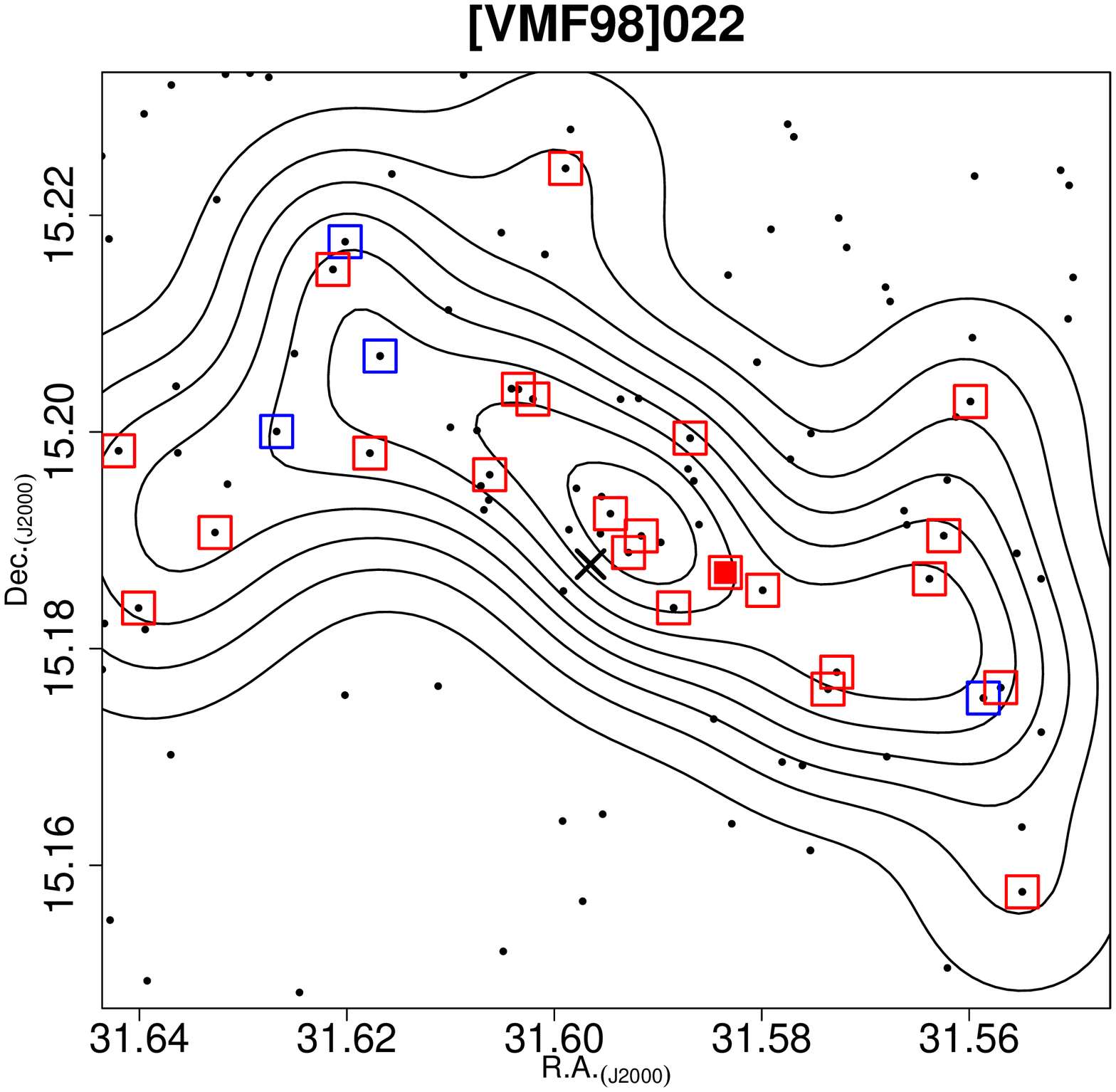}
 \includegraphics[width=80mm,height=80mm]{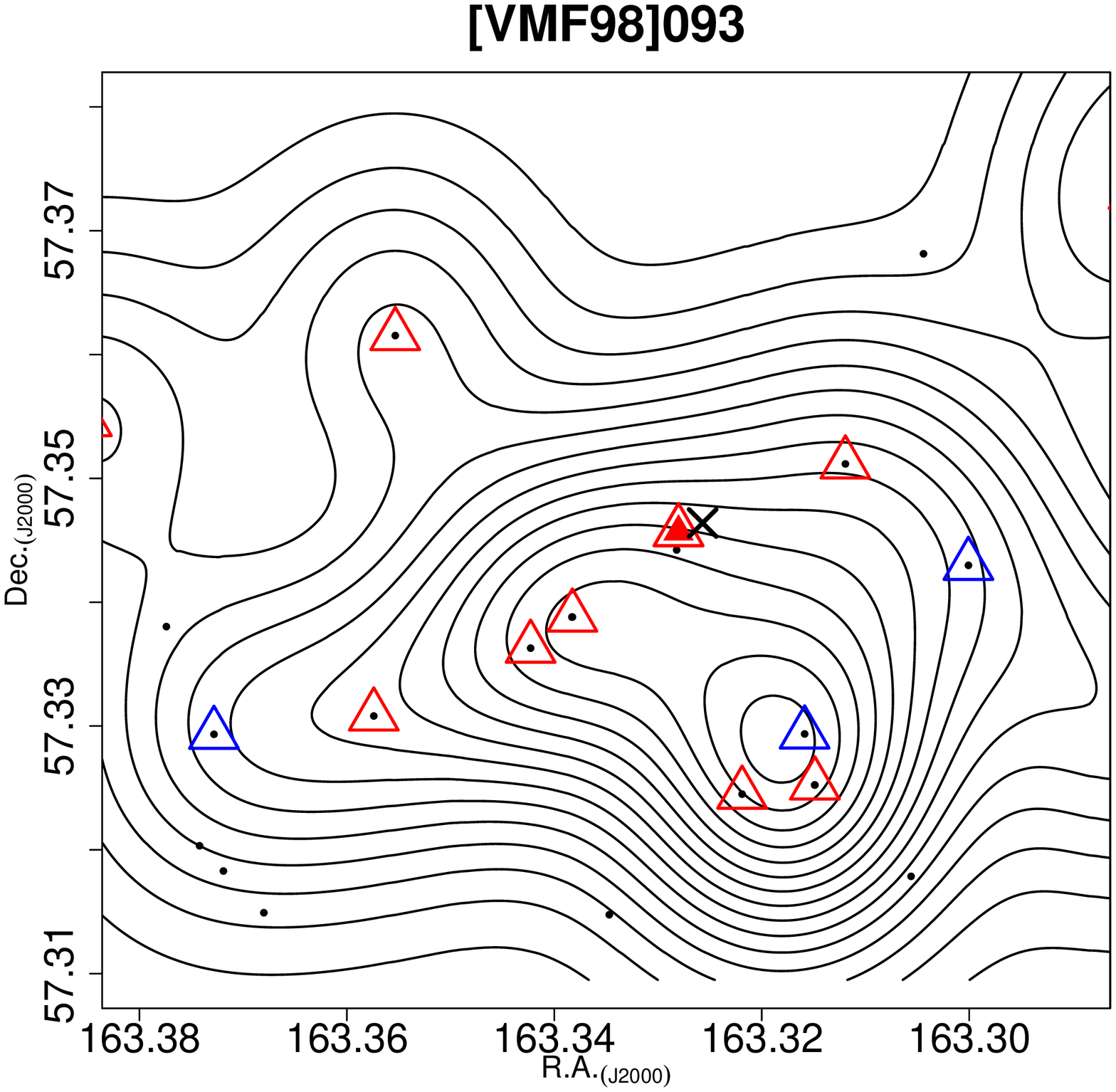}
 \includegraphics[width=80mm,height=80mm]{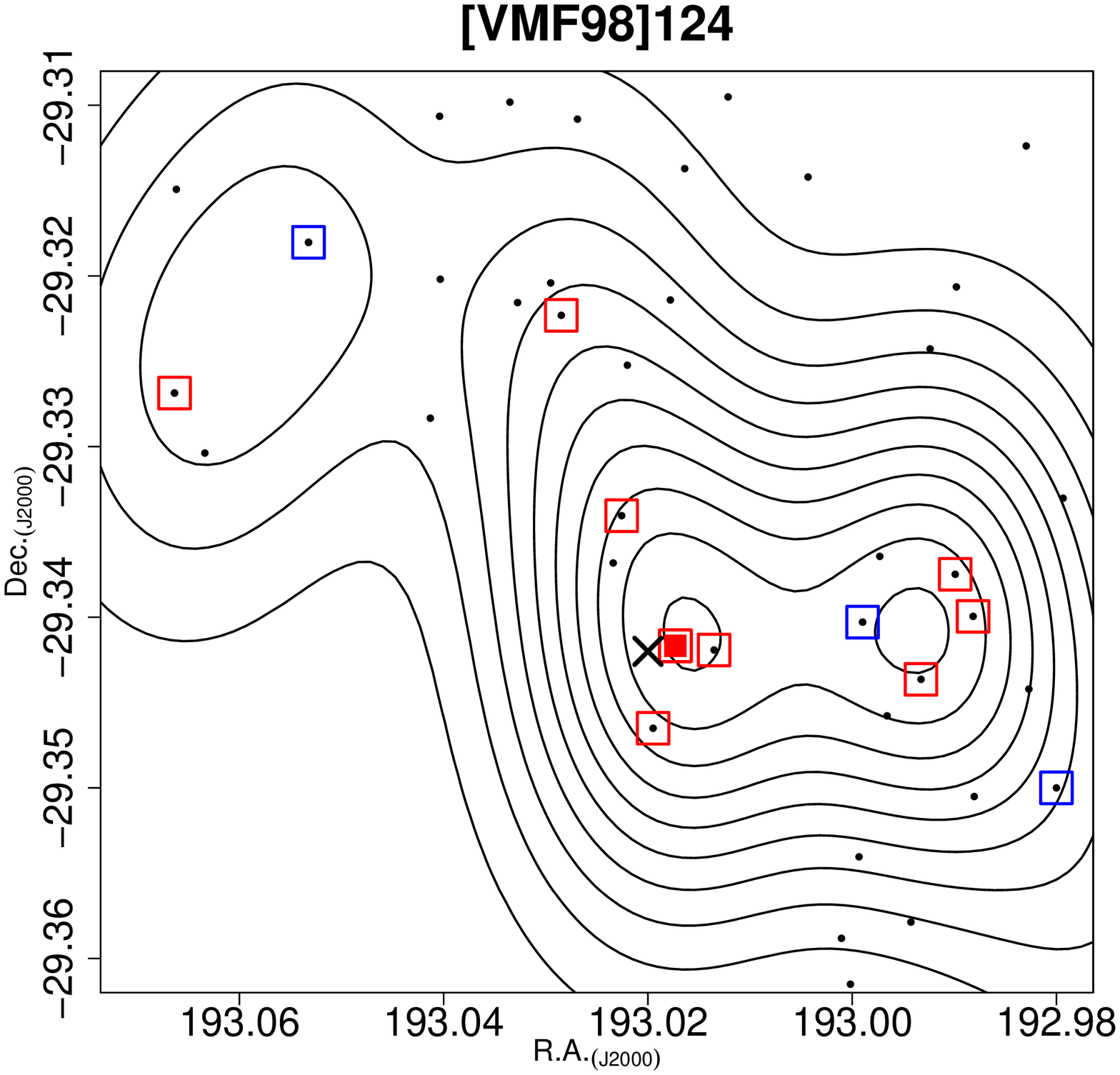}
\caption{Galaxy density maps for the low redshift galaxy clusters. 
Crosses indicate the position of the 
X-ray emission peak and filled symbols (both triangles or squares) show
the BCG locus.  Superposed to the 
 contours are the RCS and Blue samples, with the foreground 
and background galaxies being represented by black dots.}
 \label{dens_maps1}
 \end{figure*}

\begin{figure*}
 \includegraphics[width=80mm,height=80mm]{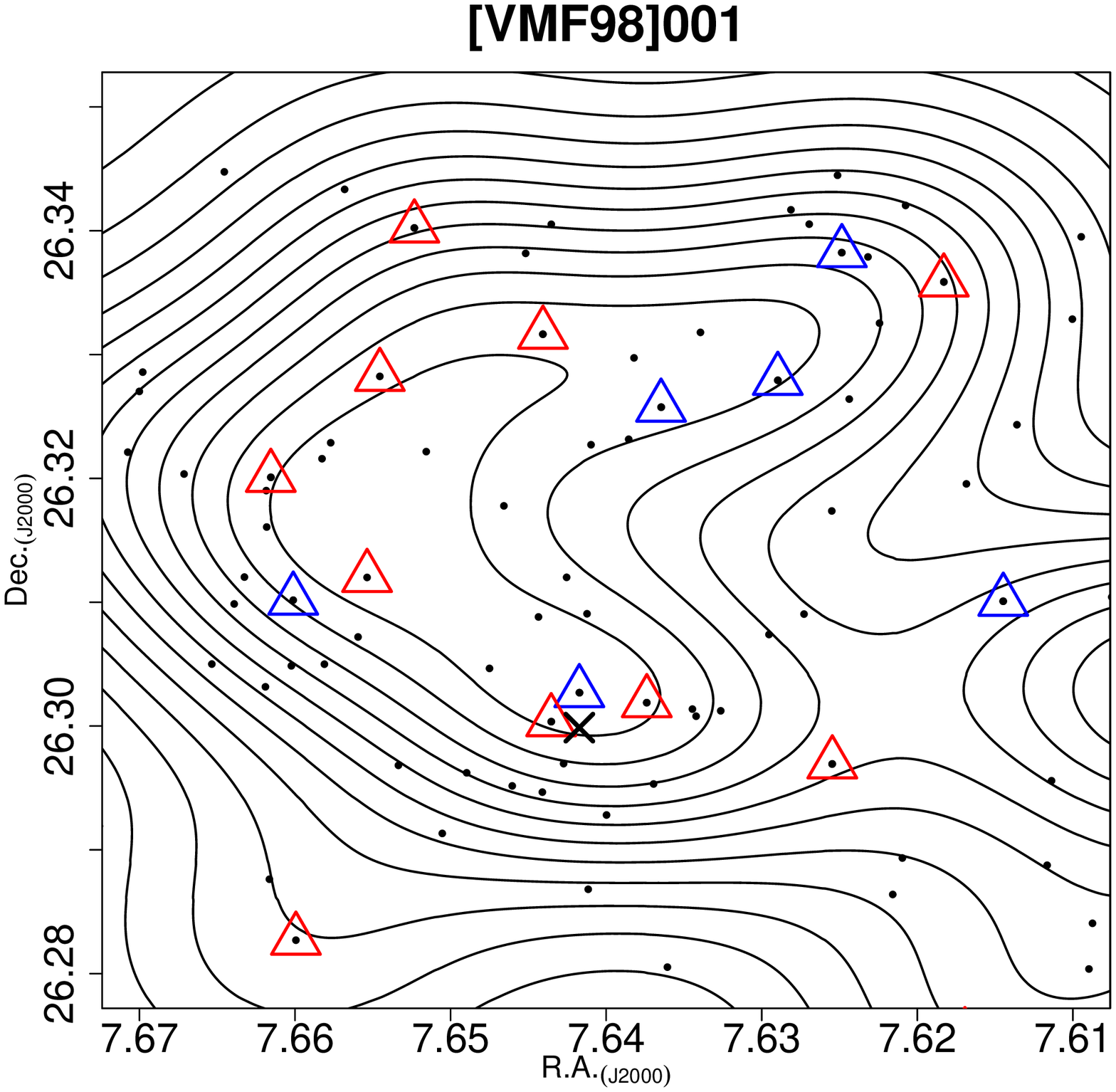}
 \includegraphics[width=80mm,height=80mm]{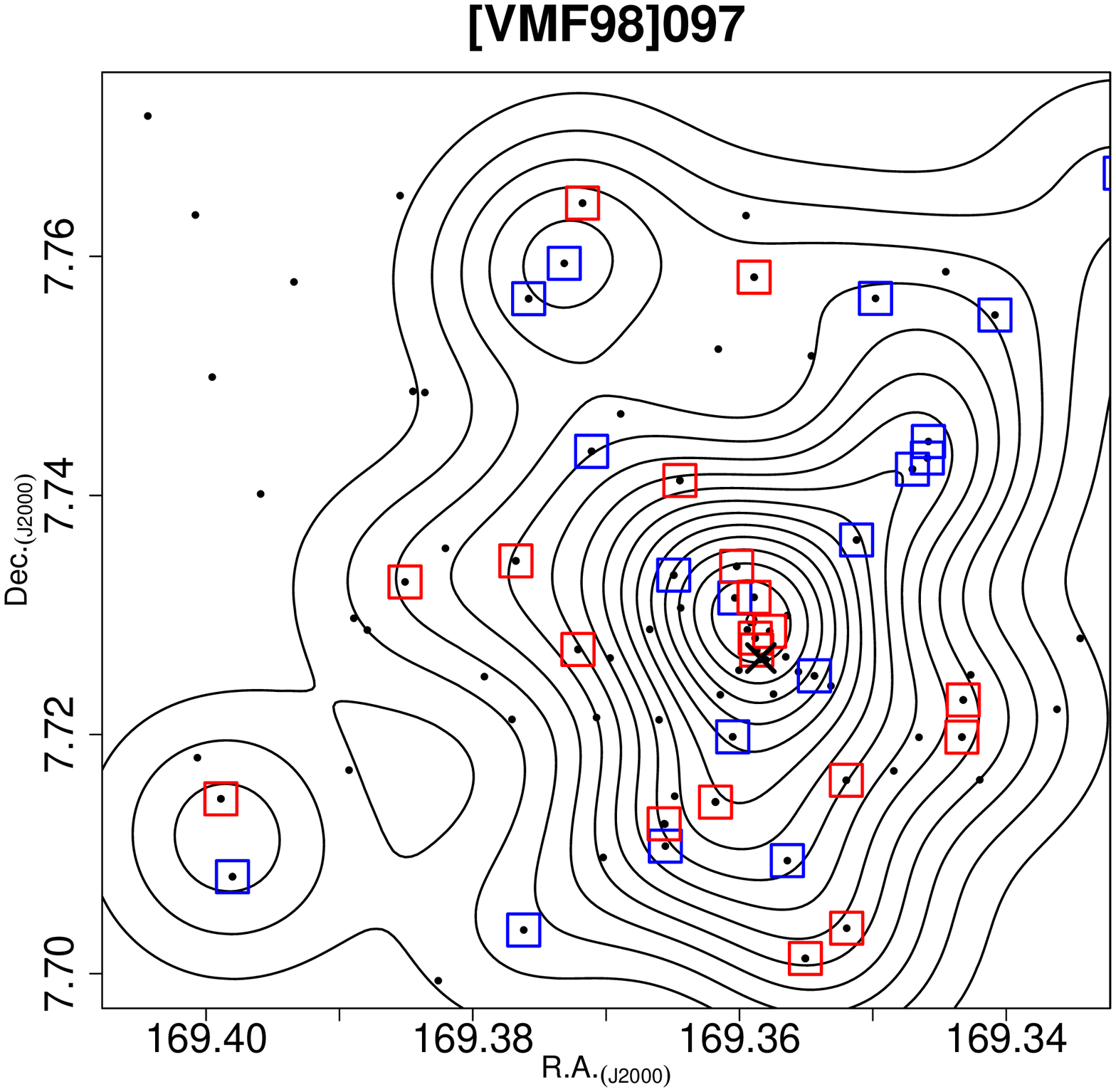}
 \includegraphics[width=80mm,height=80mm]{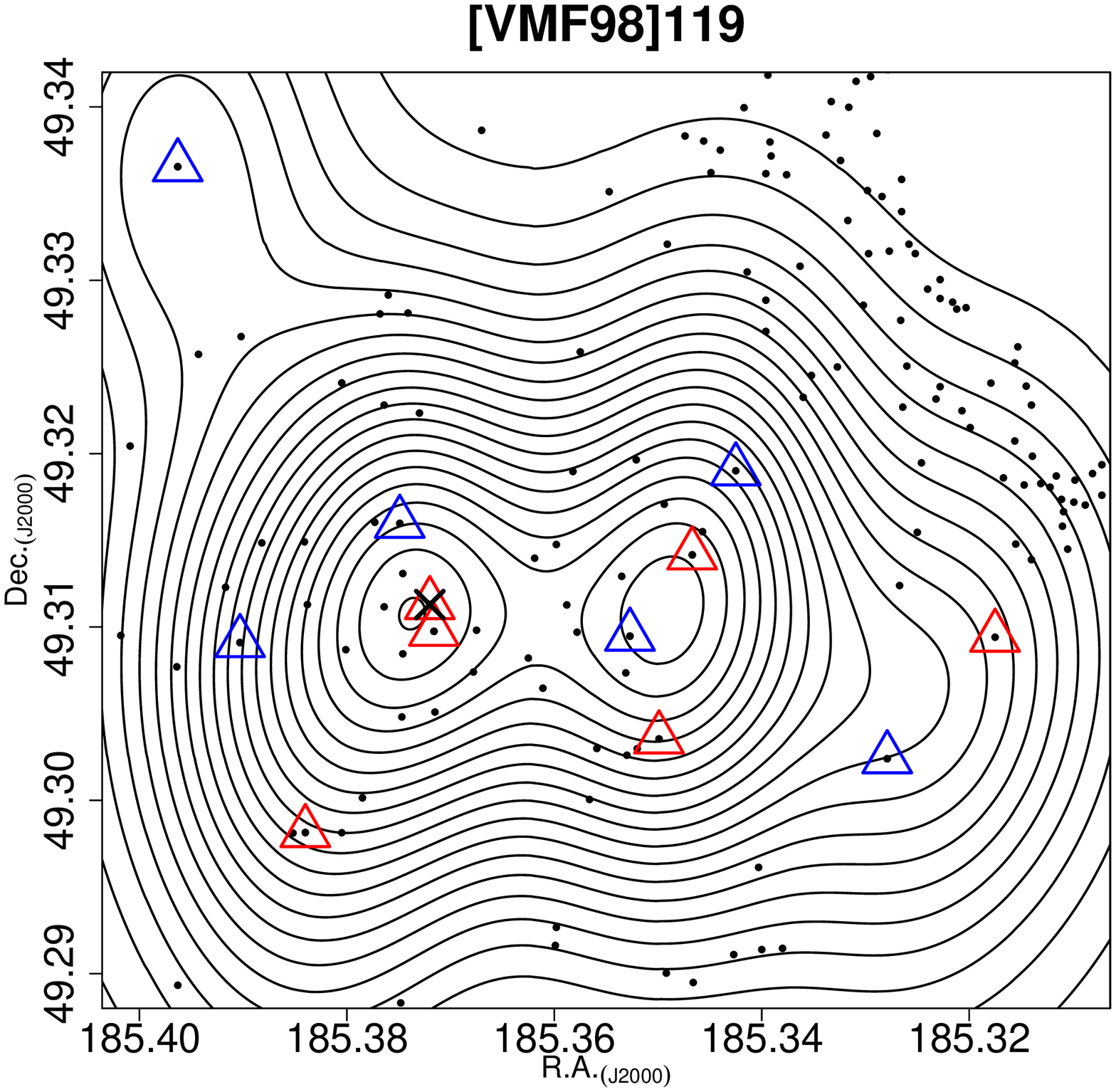}
 \includegraphics[width=80mm,height=80mm]{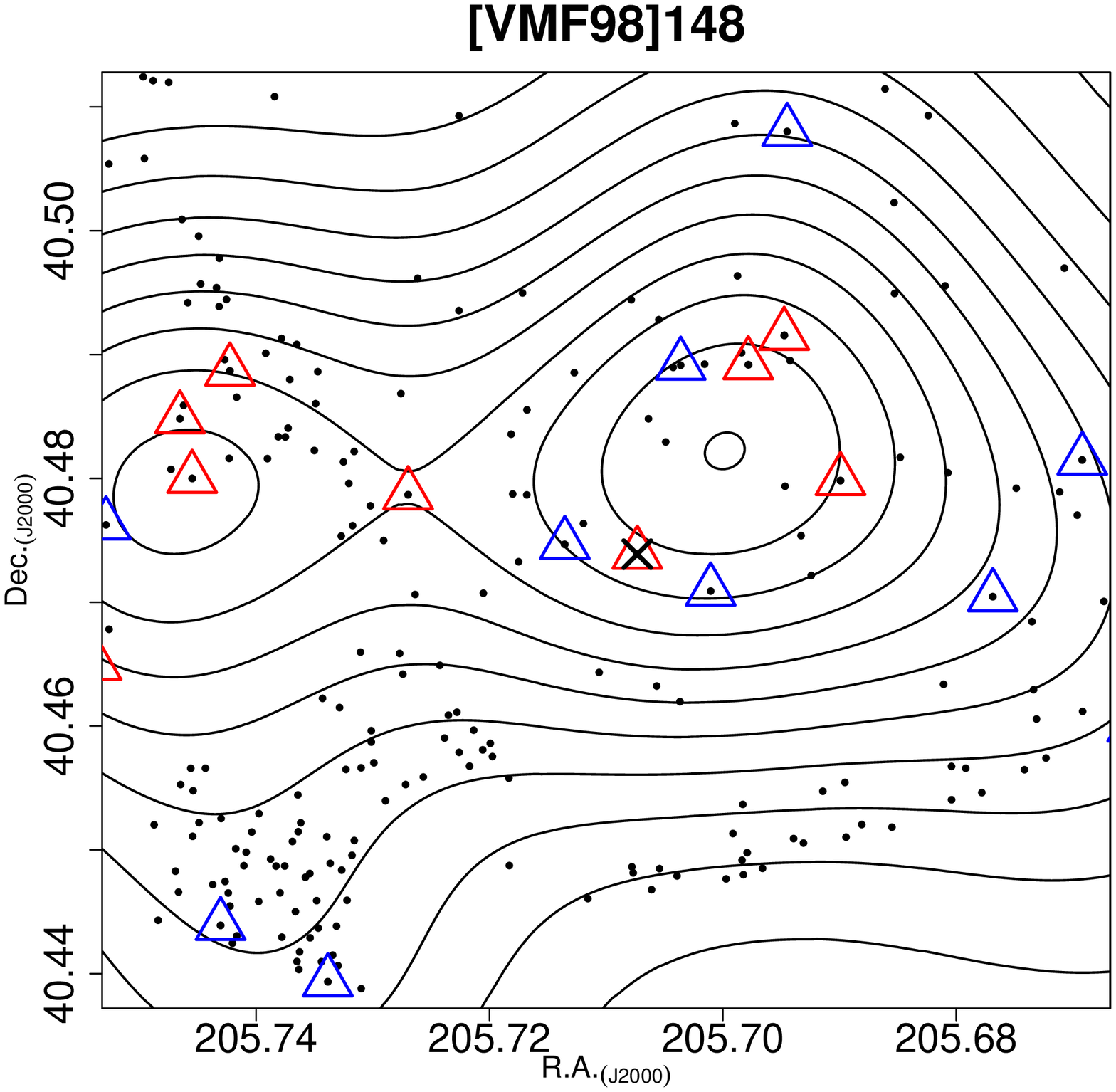}
 \caption{Galaxy density maps for the high redshift galaxy clusters. Symbols 
are similar to previous Figures.}
 \label{dens_maps2}
 \end{figure*}

 The galaxy projected distribution was generated using
 the lattice package of the R software environment\footnote{R is a 
 language and environment for statistical computing and graphics, designed by 
 Robert Gentleman and Ross Ihaka of the Statistics Department of the University 
 of Auckland. http://www.r-project.org/}, with Figures~\ref{dens_maps1} and 
~\ref{dens_maps2} showing the 
galaxy projected distribution for the three low redshift and four high redshift
galaxy clusters, respectively. The contour maps are based on 
 the RCS and Blue samples using open triangles and squares with 
the same symbol convention used
 in previous Figures. In addition, crosses indicate the position of the X-ray 
emission peak, filled triangles and squares show the BCG locus, with
black dots corresponding to foreground 
and background galaxies.  In the Figures, 
 North is up and East is to the left.

 The Figures reveal differences in the nature of the galaxy distribution.  
At lower 
 redshifts, the cluster [VMF98]022 was centrally concentrated, but shows an 
elongated 
 distribution in the NE-SW direction. The central galaxy was located at
 (RA = 02$^{h}$06$^{m}$21$^{s}$, DEC = 15$^{\circ}$11$^\prime$00$^{\prime\prime}$), 
 presenting a shift of $\sim$12$^{\prime\prime}$ in the south-west direction
 with respect to the X-ray peak emission where four red galaxies were located.
The clusters [VMF98]093 and [VMF98]124 presented a smooth single distribution,
with these results also being consistent with the well defined RCS 
found for clusters [VMF98]022 and [VMF98]124 
(Figures~\ref{cmd1} and \ref{cmd}, respectively). 
 In these three lower redshift clusters, 
 the cores were dominated by red galaxies in agreement with 
 the typical morphology-density relation (Dressler et al. 1997) found for 
massive clusters.  At higher redshifts, Figure~\ref{dens_maps2} reveals that 
the cluster [VMF98]097 had a better resolution, due to 
spectroscopic redshifts, compared to the 
other three galaxy clusters with photometric redshift estimates.

Both the present-day dynamic properties of clusters and the morphology of 
their 
member galaxies are believed to be strongly linked to the cluster accretion 
history (Girardi et al. 2011, Tonnesen \& Bryan 2009). 
Relics of this process can be found in the form of substructures in the galaxy 
distribution, since they mantain a memory of past merger events 
(White, Cohn \& Smit 2010).

The Dressler-Shectman test (DST, Dressler \& Shectman, 1988) 
evaluates the substructure in the velocity kinematics of galaxy groups 
identified in clusters in projection. By
considering the number of neighbours in projection from each galaxy, 
the mean local velocity and the local velocity 
dispersion can be inferred, which can then be compared with the values for 
the whole group.  This test has been used both in studies on individual 
clusters, e.g. Boschin et al. (2006); 
Girardi et al. (2008); Barrena et al. (2011) and in samples of clusters at
different redshifts (Einasto et al. 2012; Connelly et al. 2012), as well as
being utilized in 
numerical simulations (Knebe \& Mueller, 2000; Cohn, 2012).

Connelly et al. (2012) discussed the importance of defining the
correct radial cut in order to define substructures in galaxy clusters.  
They 
found little dynamical complexity when using X-ray based r$_{200}$ cuts 
whereas at 
larger radii (1 Mpc or greater)
significantly more dynamical substructure was obtained using the DST.
On applying this test to the three clusters with spectroscopic redshifts,
we found a lack of significant 
substructure, an expected result due to the restriction of our study to the
 central, mostly virialized 0.75 Mpc region where less substructure is 
expected. 

The [VMF98]097 cluster presented at least three 
over-densities, as shown on the contour map of Figure~\ref{densityprof2}.  In 
Paper I, we reported the 
radial velocity distribution for this cluster and ruled out a
Gaussian distribution (Hou et al. 2009).  
As seen in this Figure, the cluster core of the high redshift clusters 
did not have a 
dominant galaxy population, suggesting a mix of 
galaxies, as also observed in Figures~\ref{cmd} and \ref{cmd1}.
Carrasco et al. (2007) also studied this cluster [VMF98]097 and observed 
signs of 
dynamical complexity, while the density map presented here is slightly 
different as we analysed the inner 0.75 Mpc central region. 

 \subsection{Morphological content}

 In $\S$2.1, we discussed the visual morphological classification procedure and the adopted
morphological parameters.  It was found that 97\% of the galaxies classified 
 as early-types presented C values higher than 2.5, whereas for spiral 
galaxies,  98\% 
 had values lower than 3.0, with S0/Sa galaxies presenting values between 
 2.5 $<$ C $<$ 3.6. These distributions are illustrated in Figure~\ref{histos}.
 
\begin{figure*} 
 \includegraphics[width=60mm,height=60mm]{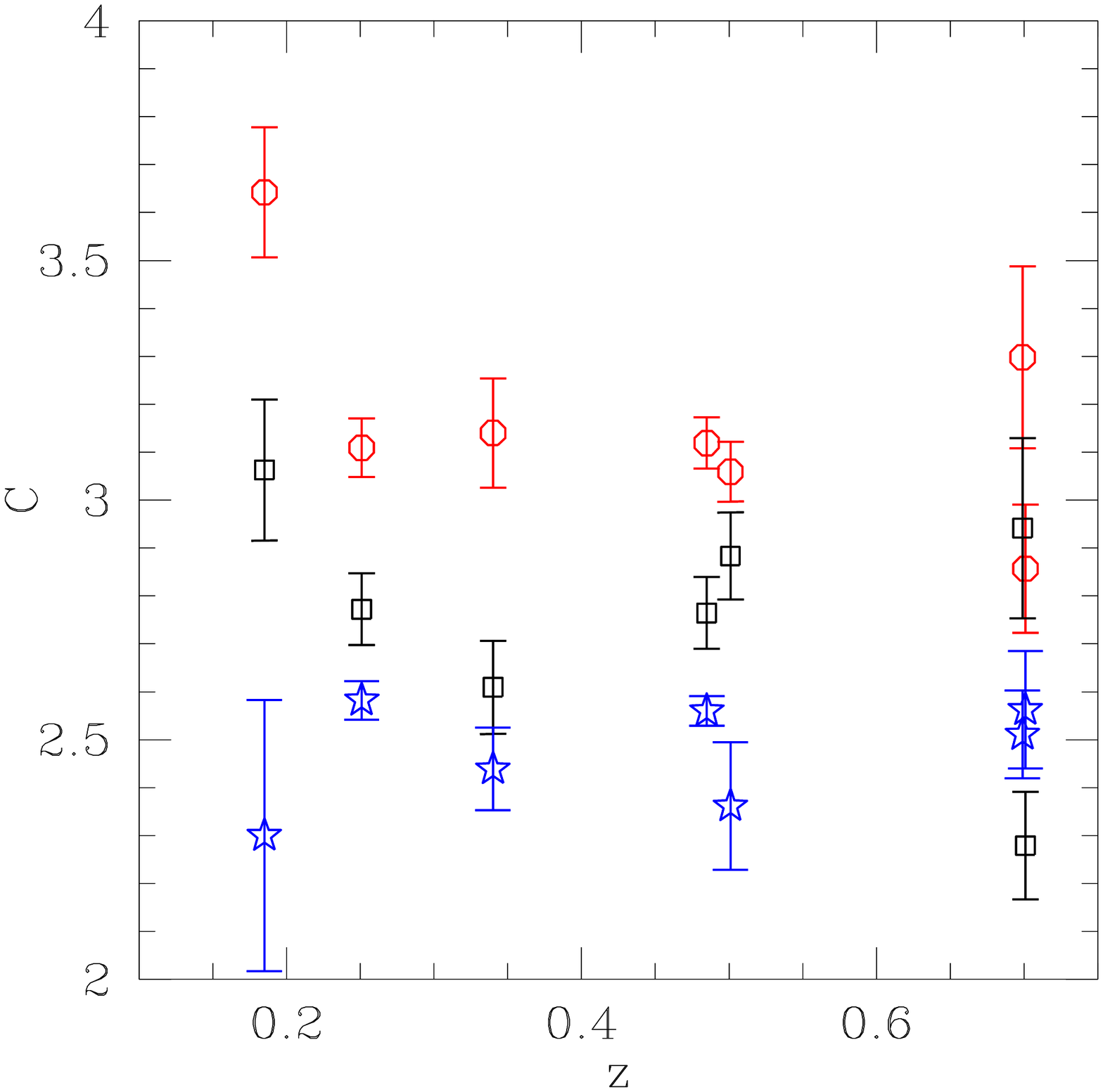}
 \includegraphics[width=60mm,height=60mm]{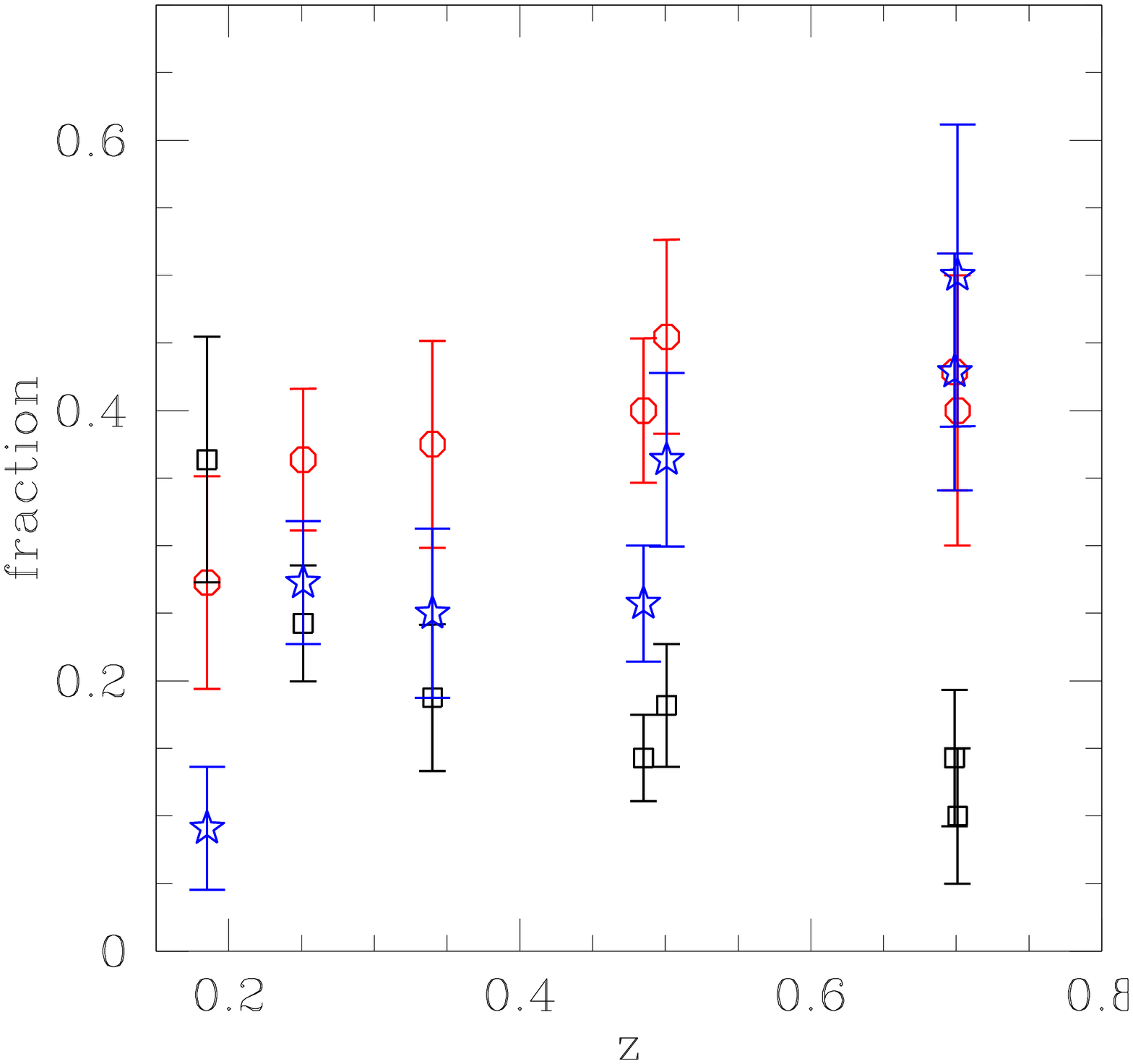}
 \caption{Morphological properties versus z for the studied galaxy clusters. 
Left panel: mean concentration 
 index for early (squares), lenticular (circles) and spiral (triangles) 
galaxies. Right panel: morphological type fraction. Error bars are based on Poisson statistics.}
 \label{morpho}
 \end{figure*}

 Figure~\ref{morpho} shows the average C values for the early, lenticular and 
spiral types
 for each galaxy cluster as a function of the cluster redshift (left panel).
  An increase in C was observed for early-type and lenticular galaxies, which
may have been related to the presence of giant galaxies in the low redshift 
clusters (C$>$4). In contrast, for late types, we found smaller C values 
 for the lower redshift cluster.  The right panel of the Figure shows 
 the fraction of these three morphological types relative to the total,
with there being a clear trend corresponding to an increase in the number of 
lenticulars at
 lower redshifts.  This effect in low-mass galaxy clusters was predicted by 
 Poggianti et al. (2009) and Dressler et al. (2009) and our data seem to
confirm this. 

 \section{SUMMARY OF THE MAIN RESULTS}

 In this paper, we have presented the $r^\prime$, $g^\prime$ and $i^\prime$ 
photometry 
 obtained with GMOS-IMAGE at Gemini North and South telescopes 
 for seven low X-ray galaxy clusters: [VMF98]001, [VMF98]022, [VMF98]093, 
[VMF98]097, [VMF98]119, [VMF98]124 and 
 [VMF98]148 in the redshift range of 
 0.18 to 0.70.  We have reduced the images and created catalogues with 
the detected objects being classified as extended following our criteria discussed in 
section $\S$2.  These catalogues contain the object coordinates, optical 
properties and
 redshift.  In addition, magnitudes, colours and morphological parameters 
such as the concentration index, ellipticity and morphological type estimates 
are also included in the 
photometric properties.  This work is the second in a series of papers aimed 
at understanding the processes involved in the 
formation and evolution of low X-ray luminosity galaxy clusters at intermediate 
redshifts.    

The RCS and the Blue samples were defined using the
 cluster members as described in $\S$3.2 and in Paper I.  In our analysis, we 
obtained the galaxy
number counts and the CMRs of the clusters, including
 the RCS, galaxy colour distributions and the relative fraction of the 
blue galaxies.  
 The galaxy radial density profiles and galaxy projected 
 distributions were also obtained for our cluster sample.

 At lower redshifts (z $<$ 0.4), the main results presented in this paper can be summarized as follows:

 \begin{itemize}

 \item The presence of a well defined RCS was found with an 
 extension of at least 4 magnitudes, characteristic of virialized 
 systems with a dominant elliptical 
 galaxy.  We also encountered some galaxies with extreme colours around
(g - r)$^{\prime}$ $\sim$ 0.8 in [VMF98]124 and 
 (r - i)$^{\prime}$ $\sim$ 0.37 in [VMF98]093, which possibly indicates
 strong star formation related to cluster dynamics.  

 \item The clusters are dominated 
by a red population, with a peak at (g - r)$^{\prime}$ $\sim$ 1.33 for 
 [VMF98]124 and (r - i)$^{\prime}$ $\sim$ 0.64 and 0.76 for  
 [VMF]022 and [VMF98]093, respectively. 

 \item The galaxy density profiles show a central concentration for 
 the RCS sample, with an absence of blue galaxies in the central region. 
Nearly 70\% of the red galaxies 
are located at the cluster core (r$_c < $200 kpc). 

 \end{itemize}

 At higher redshifts (0.4 $<$ z $<$ 0.7), the galaxy clusters studied 
revealed:

 \begin{itemize}

 \item A less important RCS with a smaller extension, but with a 
 clear presence of the blue population.  

 \item A broader galaxy colour distribution with 
 two peaks: the RCS and the blue population. A Gaussian function provided a 
good fit for both populations. 

 \item The total galaxy radial density profiles were well fitted by a single 
power law.  

 \end{itemize}

For the three galaxy clusters with spectroscopic measurements, 
no significant
detection of substructures through the DST analysis was found.

The fractions of elliptical, lenticular and blue galaxies were calculated as 
a function of cluster redshift (0.17 $<$ z $<$ 0.70), confirming an increasing 
fraction of 
blue galaxies from 0.1 to 0.5 with redshift, a fact related to 
the Butcher-Oemler effect.
As redshift increases, the fraction of lenticulars decreases from 
about 0.4 to 0.1, whereas the fraction of early-types remaines almost 
constant at about 0.35.  In addition, the concentration index of cluster 
galaxies showed no 
strong dependence on redshift, except for [VMF98]124.

The seven galaxy clusters studied in this paper is only a fraction of a larger 
sample that will be addressed in forthcoming papers. In this sense, the trends 
reported for this small cluster sample are not sufficiently robust, so 
in future work we will attempt to consolidate these findings. 

\vskip 0.5cm

{\bf ACKNOWLEDGEMENTS}
\vskip 0.5cm

Jos\'e Nilo Castell\'on acknowledges the financial support from Consejo Nacional de Investigaciones Cient\'ificas y T\'ecnicas de la Rep\'ublica Argentina (CONICET).  This work was partially supported by
Consejo de Investigaciones Cient\'ificas y T\'ecnicas (CONICET), Secretar\'ia
de Ciencia y T\'ecnica (Secyt) of Universidad Nacional de C\'ordoba and
Ministerio de Ciencia y Tecnolog\'ia (MINCyT, C\'ordoba). We thank Dr. Paul 
Hobson, native speaker, for revision of the manuscript.

SDSS-III is managed by the Astrophysical Research Consortium for the Participating Institutions of the SDSS-III Collaboration including the University of Arizona, the Brazilian Participation Group, Brookhaven National Laboratory, University of Cambridge, University of Florida, the French Participation Group, the German Participation Group, the Instituto de Astrof\'isica de Canarias, the Michigan State/Notre Dame/JINA Participation Group, Johns Hopkins University, Lawrence Berkeley National Laboratory, Max Planck Institute for Astrophysics, New Mexico State University, New York University, Ohio State University, Pennsylvania State University, University of Portsmouth, Princeton University, the Spanish Participation Group, University of Tokyo, University of Utah, Vanderbilt University, University of Virginia, University of Washington, and Yale University.

 %%%%%%%%%%%%%%%%%%%%%%%%%tables%%%%%%%%%%%%%%%%%%%%%%%%%%%%%%%%%%%%%%%%%%

 \begin{table*}
 \caption{Low X-ray luminosity Galaxy Clusters}
   \label{table1}
 {
  \begin{tabular}{ccrrclccc}
 \hline
 \hline
  [VMF98]       & Right Ascension &   Declination      &       $L_X$       &z & Program & g$^{\prime}$ & r$^{\prime}$ & i$^{\prime}$\\
   Id. &(J2000)          &         (J2000)    &   [10$^{43}$ cgs]  & & Id. & &              &             \\ 

 \hline
      &         &  &                     &              &      \\
 
  001 & 00 30 33.2  & +26 18 19  & 26.1	& 0.500 & GN-2010B-Q-73 & -- & 15$\times$300 & 15$\times$150  \\
  022 & 02 06 23.4  & +15 11 16  & 3.6 & 0.248 &  GN-2003B-Q-10 & -- & 4$\times$300 & 4$\times$150 \\
  093 & 10 53 18.4  & +57 20 47  & 1.4 & 0.340 &  GN-2011A-Q-75 & --  & 5$\times$600 & 4$\times$150 \\
  097 & 11 17 26.1  & +07 43 35 & 6.4 & 0.477 & GS-2003A-SV-206 & 12$\times$600 & 7$\times$900 & -- \\
  119 & 12 21 24.5  & +49 18 13  & 42.7	& 0.700 & GN-2011A-Q-75 & -- & 7$\times$190 & 4$\times$120 \\
  124 & 12 52 05.4  & -29 20 46 & 3.4   & 0.188 & GS-2003A-SV-206 & 5$\times$300  & 5$\times$600 & -- \\
  148 & 13 42 49.1 & +40 28 11  & 16.2& 0.699 & GN-2011A-Q-75 & --& 7$\times$190 & 5$\times$120 \\
               &  &                             &    &          &      \\ 

 \hline
 \hline
 \end{tabular}

 }
 \end{table*}

 \begin{table*}
 \caption{Completeness levels in the r$^{\prime}$ band.}
   \label{table2}
 {
  \begin{tabular}{ccrcccccc}
 \hline
 \hline
  [VMF98]       & 50\% &  90\% \\
   Identification & level   & level \\
 \hline
               &  &   \\ 

  001 & 23.20$\pm$0.04 & 23.70$\pm$0.05 \\
  022 & 22.00$\pm$0.09 & 23.00$\pm$0.02 \\
  093 & 22.10$\pm$0.07 & 23.00$\pm$0.20 \\
  097 & 23.10$\pm$0.02 & 24.00$\pm$0.04 \\
  119 & 23.20$\pm$0.06 & 23.70$\pm$0.10 \\
  124 & 21.50$\pm$0.04 & 23.00$\pm$0.05 \\
  148 & 23.20$\pm$0.08 & 23.60$\pm$0.16 \\ 
               &  &    \\ 
 \hline
 \hline
 \end{tabular}
 }
 \end{table*}

 \begin{table*}
 \caption{Galaxy member properties of the low X-ray luminosity cluster [VMF98]022.}
   \label{table3}
 {
  \begin{tabular}{ccrcccccc}
 \hline
 \hline
Galaxy Id.	& RA (J2000) 	&DEC (J2000)	&r$^\prime$	& (r-i)$^{\prime}$& C		& b/a		&T	& z	\\
 \hline 
                         &		&		&			& 		  & 		&		&	&		\\
J020622.6+151132	&02:06:22.698	&+15:11:32.86	&21.1415 & 0.6319	  & 2.422	&0.032		& 3	& 0.24680	\\
J020631.8+151126	&02:06:31.848	&+15:11:26.73	&22.2924 & 0.6706	  & 2.543	&0.321		& 2	& 0.24843	\\
J020614.9+151125	&02:06:14.981	&+15:11:25.59	&21.5822 & 0.5430 	  & 2.501	&0.338		& 3	& 0.25017	\\
J020621.9+151125	&02:06:21.984	&+15:11:25.54	&19.4243 & 0.7473	  & 3.246	&0.038		& 1	& 0.25100  	\\
J020622.2+151120	&02:06:22.277	&+15:11:20.00	&21.1816 & 0.6866	  & 2.802	&0.025		& 1	& 0.24878	\\
J020615.3+151111	&02:06:15.316	&+15:11:11.29	&20.7716 & 0.7061	  & 2.686	&0.273		& 2	& 0.24732	\\
J020619.1+151107	&02:06:19.175	&+15:11:07.49	&20.7354 & 0.7264	  & 2.612	&0.337		& 3	& 0.25040	\\
J020621.2+151101	&02:06:21.231	&+15:11:01.58	&17.9699 & 0.7517	  & 4.448	&0.196		& 1	& 0.24737	\\
J020633.6+151101	&02:06:33.613	&+15:11:01.54	&18.5363 & 0.7041	  & 3.607	&0.168		& 2	& 0.25032	\\
J020617.4+151040	&02:06:17.455	&+15:10:40.28	&19.9922 & 0.7202	  & 3.386	&0.072		& 1	& 0.24611	\\
J020613.6+151035	&02:06:13.667	&+15:10:35.04	&20.3462 & 0.7189	  & 2.777	&0.234		& 1	& 0.24630	\\
J020617.6+151034	&02:06:17.657	&+15:10:34.60	&21.7576 & 0.6954	  & 2.552	&0.203		& 3	& 0.24893	\\
J020614.0+151031	&02:06:14.061	&+15:10:31.70	&20.9661 & 0.7069	  & 2.781	&0.256		& 2	& 0.24870	\\
J020613.1+150927	&02:06:13.161	&+15:09:27.30	&20.2152 & 0.7201	  & 2.661	&0.242		& 2	& 0.25100  	\\
J020623.7+151327	&02:06:23.735	&+15:13:27.61	&18.7835 & 0.7052	  & 3.507	&0.301		& 1	& 0.24715	\\
J020628.8+151303	&02:06:28.836	&+15:13:03.25	&21.8539 & 0.5855	  & 2.428	&0.032		& 2	& 0.25100  	\\
J020629.1+151254	&02:06:29.119	&+15:12:54.07	&19.6068 & 0.6969	  & 2.821	&0.376		& 2	& 0.24857	\\
J020628.0+151225	&02:06:28.023	&+15:12:25.28	&21.4024 & 0.8439	  & 2.801	&0.679		& 3	& 0.24872	\\
J020624.8+151214	&02:06:24.830	&+15:12:14.23	&19.9894 & 0.6956	  & 2.930	&0.41 		& 1	& 0.24389	\\
J020624.4+151211	&02:06:24.488	&+15:12:11.01	&20.9378 & 0.7046	  & 2.651	&0.287		& 2	& 0.24554	\\
J020614.3+151210	&02:06:14.362	&+15:12:10.22	&20.7151 & 0.6902	  & 2.535	&0.382		& 3	& 0.24854	\\
J020630.4+151200	&02:06:30.424	&+15:12:00.25	&19.1867 & 0.7230 	  & 3.315	&0.195		& 1	& 0.25001  	\\
J020620.8+151158	&02:06:20.852	&+15:11:58.04	&19.9817 & 0.7054	  & 3.113	&0.39 		& 1	& 0.24899	\\
J020634.0+151153	&02:06:34.072	&+15:11:53.80	&20.0432 & 0.6971	  & 2.739	&0.302		& 1	& 0.24901	\\
J020628.2+151153	&02:06:28.258	&+15:11:53.03	&20.5783 & 0.7432	  & 2.735	&0.347		& 1	& 0.24830	\\
J020625.4+151145	&02:06:25.488	&+15:11:45.94	&21.3265 & 0.6721	  & 2.875	&0.092		& 5	& 0.24600	\\
                         &		&		&			& 		  & 		&		&	&		\\
 \hline                                                                                                          
 \hline
 \end{tabular}
 }
 \end{table*}

 \begin{table*}
         \caption{Red Cluster Sequence: Linear Regressions}
   \label{table4}
 {\small
                 \begin{tabular}{crll}
                 \hline
                 \hline                                    
                 [VMF98]	&Number of	&Slope	        	&Zeropoint           \\
Id.               &Galaxies	        &	        	        &	                   \\
                 \hline   
 & & & \\	 	 	 
001 & 13 & -0.021$\pm$0.006 & 2.537$\pm$0.011 \\
022 & 21 & -0.023$\pm$0.099 & 0.065$\pm$0.011 \\
093 & 10 & -0.011$\pm$0.017 & 0.084$\pm$0.010 \\
097 & 19 & -0.029$\pm$0.008 & 1.932$\pm$0.012 \\
119 &  8 & -0.019$\pm$0.006 & 3.083$\pm$0.016 \\
124 & 10 & -0.025$\pm$0.013 & 1.328$\pm$0.019 \\
148 & 10 & -0.030$\pm$0.011 & 3.335$\pm$0.019 \\	 	 	
                  \hline                                                        
                  \hline
         \end{tabular}
         }
 \end{table*}

 \begin{table*}
         \caption{Gaussian fits to the galaxy populations: Statistical values}
   \label{table5}
         {\small
                 \begin{tabular}{ccccccc}
                 \hline
                 \hline 
  [VMF98]  & Red & Population & & Blue & Population & \\
  Id.      & Mean & Standard & $\chi$$^2$ & Mean & Standard & $\chi$$^2$ \\ 
           &      & Deviation &           &      & Deviation & \\
 \hline
         & &           & & &	       & \\
001	& 2.080	& 0.051	& 0.040	& 1.752 & 0.167	& 0.545	\\
022	& 0.647	& 0.025	& 0.599	& - & - & - \\
093	& 0.760	& 0.043	& 0.396	& - & - & - \\
097	& 1.941 & 0.037	& 0.736	& 1.463 & 0.311	& 1.083	\\
119	& 2.686	& 0.033	& 0.598	& 2.134 & 0.258	& 0.280	\\
124	& 1.328	& 0.058 & 0.292	& - & - & - \\
148	& 2.678 & 0.064	& 0.280	& 2.091 & 0.264	& 0.600 \\
                 \hline                                                        
                 \hline                                                        
         \end{tabular}
         }
 \end{table*}

 \begin{table*}
         \caption{The $\alpha$ parameter of the power law fit.}
   \label{table6}
         {\small
                 \begin{tabular}{ccc}
                 \hline
                 \hline 
  [VMF98]  &  $\alpha$ & rms \\
  Id.      &         &     \\
 \hline
           &         &     \\
001	& -1.371$\pm$0.285 & 2.323$\pm$0.473  \\
022	& -0.519$\pm$0.101 & 1.785$\pm$0.350  \\
093	& -0.608$\pm$0.162 & 2.001$\pm$0.535  \\
097	& -0.908$\pm$0.163 & 1.700$\pm$0.305  \\
124	& -0.945$\pm$0.272 & 2.020$\pm$0.583  \\
119	& -1.212$\pm$0.323 & 2.299$\pm$0.614  \\
148     & -1.182$\pm$0.264 & 2.281$\pm$0.510  \\
                 \hline                                                        
                 \hline                                                        
         \end{tabular}
         }
 \end{table*}

\clearpage

\end{document}